\documentclass[12pt,preprint]{aastex}
\usepackage{graphics,epsfig,lscape}
\pdfoutput=1
\makeatletter

\slugcomment{}
\shorttitle{Searching for compact objects in  SNRs G27.8+0.6 and G28.8+1.5 }
\shortauthors{Misanovic, Kargaltsev \& Pavlov}

\makeatother

\begin{document}

\title{Searching for compact objects in SNRs G27.8+0.6 and G28.8+1.5\footnote{Based on observations obtained with XMM-Newton, an ESA science mission with instruments and contributions directly funded by ESA Member States and NASA.} }

\author{Zdenka Misanovic$^1$, Oleg Kargaltsev$^2$ and George G.\ Pavlov$^3$}

\affil{$^1$School of Physics, Monash University, Melbourne, 3800 VIC,
  Australia\\
$^2$Dept. of Astronomy, University of Florida, Gainesville, FL
32611-2055, USA \\  
$^3$Dept.\ of Astronomy and Astrophysics,
The Pennsylvania State University, \\
525 Davey Lab., University Park,
PA 16802}

\begin{abstract}
We analyzed {\sl XMM-Newton} observations of two center-filled supernova remnants, G27.8+0.6
 and G28.8+1.5, to search for pulsars/pulsar wind nebulae (PWNe) and other types of neutron stars associated with these remnants. We discovered a PWN candidate within the extent of the centrally-peaked radio emission of G27.8+0.6. The X-ray morphology of the PWN candidate and its
 position with respect to the host supernova remnant suggest that the alleged pulsar might be moving away from the supernova remnant's center with a transverse velocity of around 100--200 km~s$^{-1}$. The majority   of the detected  X-ray point sources in both fields are classified as main-sequence stars,  based on their bright optical counterparts and relatively soft X-ray spectra. The remaining medium-hard and hard sources are most probably either AGNs or cataclysmic variables, although we cannot completely rule out the possibility that some of these sources are neutron stars.

\end{abstract}


\section{Introduction}
\label{introduction}

 Stellar evolution models predict that the  core-collapse supernova
 explosions  produce both a supernova remnant (SNR) and
 either a
  neutron star (NS) or a black hole (BH). The exact fraction of these
 two types of compact objects
 is not well constrained, but considering that the majority of  SN
 explosions are believed to be the core-collapse events,
 a significant fraction of SNRs are expected to harbour NSs.   
A significant number of these NSs are rapidly spinning, and are
 often detected as radio, X-ray or $\gamma$-ray pulsars.

A young energetic pulsar
 loses
most of its rotational energy in form of a pulsar wind 
 \citep{1974MNRAS.167....1R}, which interacts with the surrounding medium
 and forms 
 a pulsar wind nebula (PWN) that is observed as radio and/or 
X-ray  diffuse emission  \citep{2006ARA&A..44...17G}.
Due to the complexity of this interaction and anisotropy of the pulsar wind,
 PWNe display various
 shapes including  toroidal structures, jets, and sometimes 
 cometary-shaped
tails formed when the pulsar is moving with a supersonic
speed through the ambient medium \citep{2008AIPC..983..171K}.
Hence, to obtain a better census of the NS--SNR associations,
 high-resolution X-ray
(and radio) observations are very useful, as they can  
 identify PWNe.

Recent sensitive, high-resolution X- and  $\gamma$-ray 
 observations have also
 revealed
  unexpected diversity in the populations of NSs
 associated with SNRs, which, in addition to classical radio pulsars, also include anomalous X-ray pulsars (AXPs), soft gamma-ray repeaters (SGRs), compact central objects (CCOs),
 and perhaps other classes of radio-quiet NSs 
 \citep{2002ASPC..271....3K}. A class of radio-quiet NSs was predicted
  by \citet{1980PASP...92..125S}, who
 suggested that due to their strong magnetic fields, these NSs will decelerate
 rapidly and become inactive in several hundreds of years after the SN
 explosion. Alternatively, NSs can be born with low magnetic fields \citep[``antimagnetars''; e.g., see][and references therein]{2010ApJ...709..436H}.  
 The CCO in Cas\,A,
 revealed in
 the first-light {\sl Chandra} image \citep[see also][]{2009ApJ...703..910P}, is an example of such a source.

Out of  almost 50  NSs possibly associated with SNRs
\citep[listed in Table 1 of ][]{2002ASPC..271..233K},
 $\sim$30\% were detected as
 radio pulsars, $\sim$30\% as various radio-quiet compact objects
 (AXPs, SGRs,
 CCOs), while the remaining NS--SNR associations were proposed because a
 diffuse, non-thermal X-ray nebula (PWN), and/or radio synchrotron nebula
 (plerion) were detected within the SNR, suggesting the presence of a
 pulsar. \citet{2008AIPC..983..171K} list a total of 18 SNR-PWN-pulsar
 associations and 14 PWNe in which no pulsar has been detected to date.
 \citet{2004ApJS..153..269K2006ApJS..163..344K,2006ApJS..163..344K}
 searched for 
NSs in a volume-limited (closer than 5\,kpc)  sample of radio-shell  SNRs,
 but found no
NSs down to the luminosity limit of typical NSs of the same
ages as the observed SNRs.
 To  search for  compact stellar remnants and/or PWNe in all
 center-filled and composite SNRs  within
 5 kpc, a snapshot survey has been carried out with {\sl XMM-Newton}.
 We present the analysis of  
two such SNRs, G27.8+0.6 and G28.8+1.5.

The center-filled SNR G27.8+0.6, with a relatively low surface brightness,
 was
 detected in single-dish 
 pointed observations
at 2.69, 4.75 and 10.2 GHz \citep{1984A&A...133L...4R}.
The central part is strongly polarized, indicating a non-thermal origin and
supporting the SNR classification.  
\citet{1984A&A...133L...4R} estimated the age and  distance of the SNR
to be 35--55 kyrs and 2--3 kpc, respectively, adopting the evolutionary
 model for center-filled SNRs of \citet{1980A&A....90..269W}.

G28.8+1.5 was detected by \citet{1994A&A...286L..47S} in {\sl ROSAT}
 observations as 
a diffuse X-ray emission covering a region of $80'\times50'$, and
 identified by non-thermal radio emission as a $\sim$30-kyr-old  SNR at a
 distance of $\sim$4 kpc.
 \citet{1994A&A...286L..47S}  also reported a 5.45 s period
 pulsations from
a point source (which is not in the {\it XMM-Newton}
field of view) in the vicinity of this SNR.  However, this result has not been
confirmed later \citep{2000PASJ...52..181S}.

In a search for compact objects in  these two center-filled SNRs,
 we  analyzed
all X-ray sources detected by {\it XMM-Newton} in their fields.
To classify the detected objects and select PWN/NS candidates, we used the inferred X-ray
properties and cross-correlation with optical, infrared, and radio data. 
In Section~\ref{observations} we summarize our
data  analysis and present the compiled catalogs of the detected X-ray
 sources
 in both
 fields (Section~\ref{catalogues}), and also a  search for their counterparts
 at
 other wavelengths and the
 proposed classifications (Section~\ref{class}).
Finally, we discuss  PWN/NS candidates in 
 Section~\ref{PWNcandidates}.

\begin{figure}
\includegraphics[height=6.2cm,angle=0]{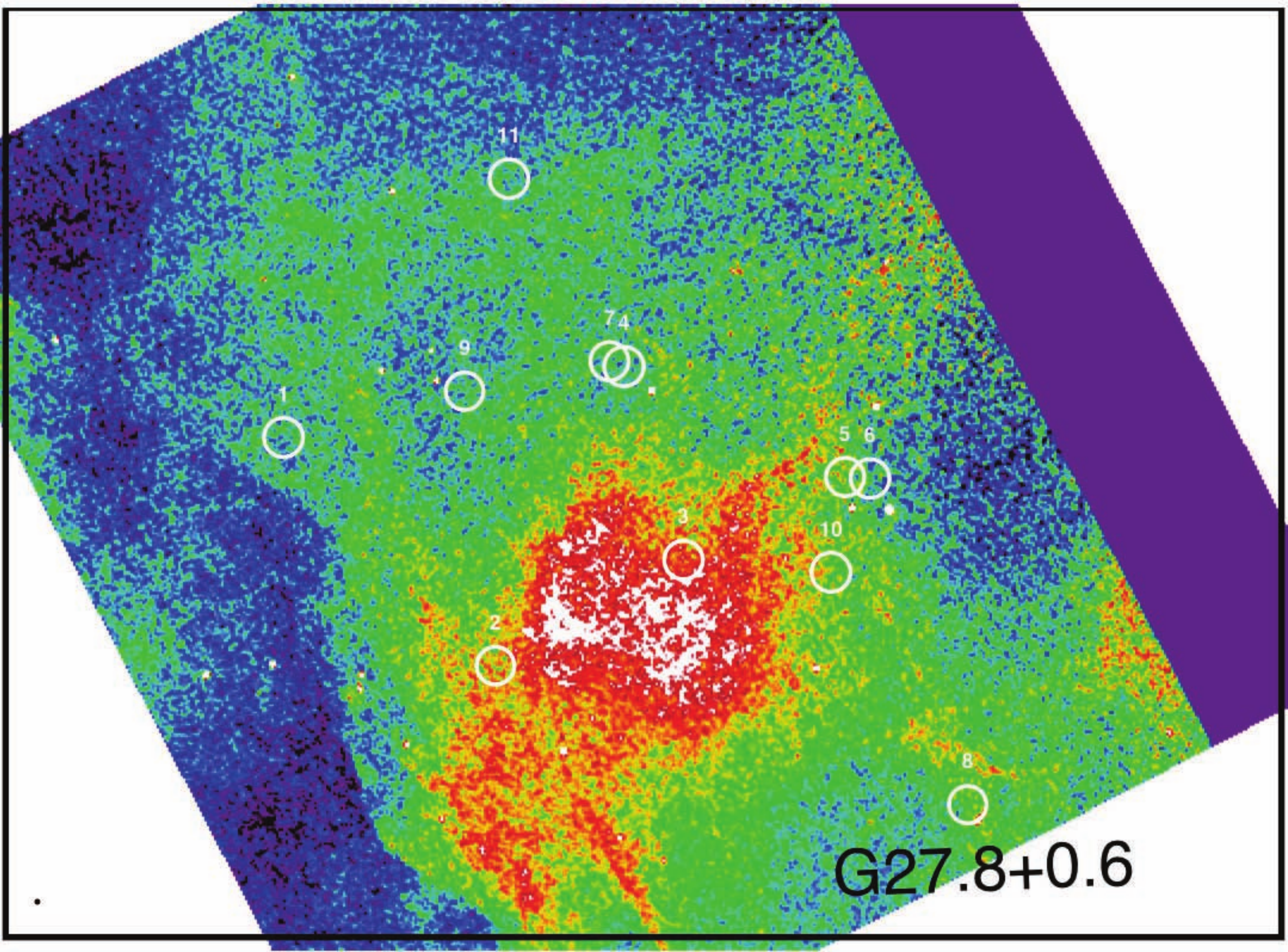}
\includegraphics[height=6.2cm,angle=0]{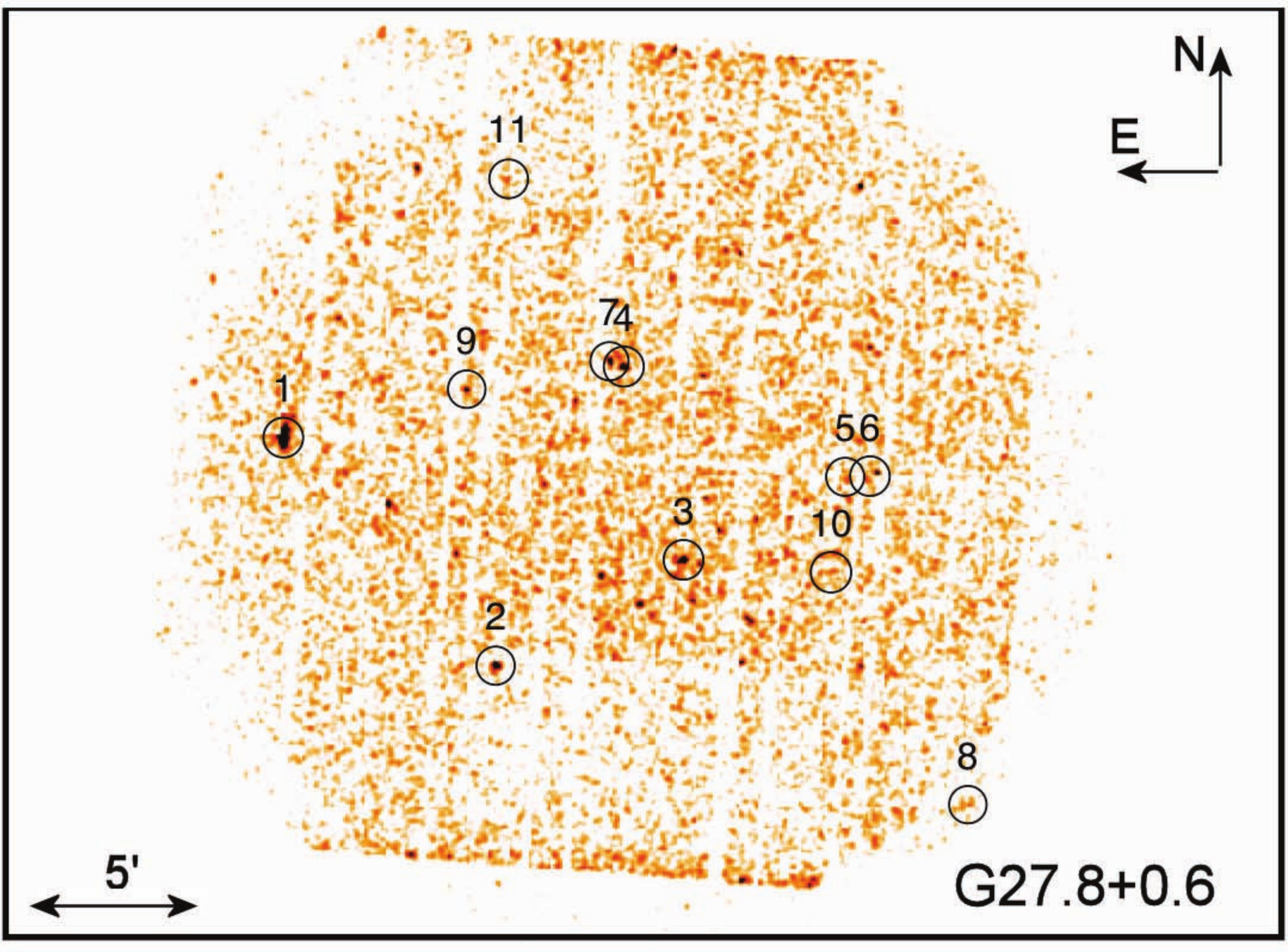}
\includegraphics[height=6.2cm,angle=0]{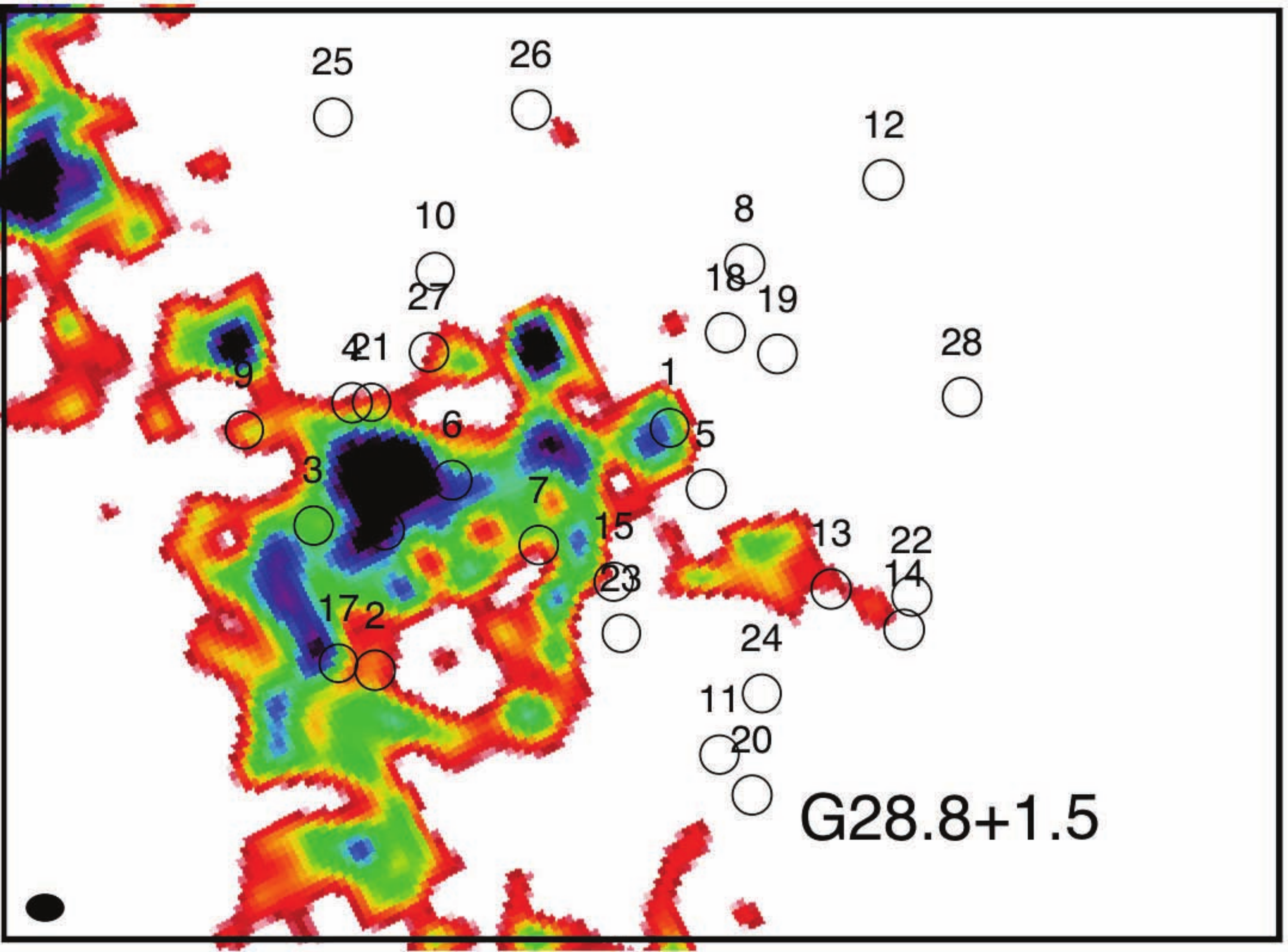}
\includegraphics[height=6.2cm,angle=0]{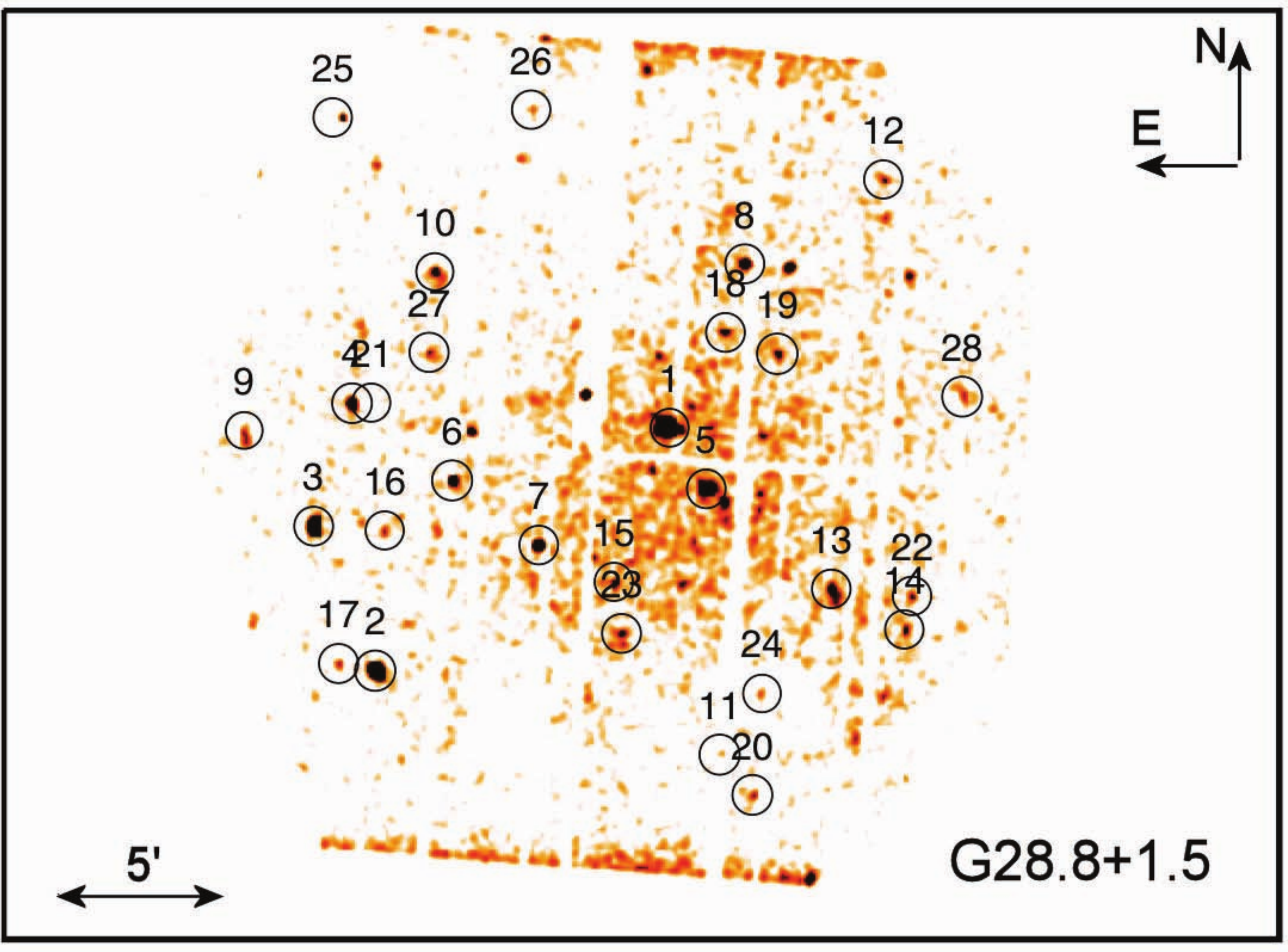}
\caption{{\bf Left:} VLA images of G27.8+0.6 at 1.4 GHz ({\it top}) and 
  G28.8+1.5 at 0.325 GHz ({\it bottom}). Both images are
extracted from
  MAGPIS \citep{2005AJ....130..586W}. The radio beams ($\sim 6''\times 5''$ for G27.8+0.6 and $\sim 66''\times 45''$ for G28.8+1.5) 
  are shown in the lower left corner of each image. 
{\bf Right:}  The broad band (0.2-12.0 keV) gray scale {\sl XMM-Newton} EPIC
 images of the
 observations 0301880401  ({\it  top}) and 0301880901
  ({\it bottom}), 
 combining PN, MOS1,
and MOS2 data. The images are spatially aligned with the corresponding radio images, and smoothed with a Gaussian of FWHM 5\arcsec.
The intensity scale is linear from 0 to 2.6943 ct s$^{-1}$ pixel$^{-1}$.
The pixel size is 4\arcsec. The circles (radius 36$''$) represent the sources
 detected
above the likelihood threshold of 7. 
The numbers correspond to the SRC IDs from Tables~\ref{cat1} to \ref{opt2}. 
 \label{radio-xmm1} }
\end{figure} 

.

\section{Observations and results}
\label{observations}
 G27.8+0.6 and G28.8+1.5 were observed in the Full
Frame mode using medium filters.
Table~\ref{obslog} lists  the observation IDs and dates, the total exposure times and the 
 exposure times after high-background screening.
The data analysis was performed using the {\sl XMM-Newton} Science Analysis System
(SAS) v7.1.0.  
We produced PN, MOS1, and MOS2 event files
 using
the SAS tasks {\tt epchain} and 
  {\tt emchain}, and then excluded the times with high background using a
  threshold count-rate of 1.5 cts~s$^{-1}$ for all three detectors.

To search for X-ray sources in the two SNR fields, we  produced PN, MOS1 and
MOS2 images 
 in five
 energy bands:     
 0.2--0.5 keV, 0.5--1.0 keV,
1.0--2.0 keV, 2.0--4.5 keV, and 4.5--12.0 keV (bands B1 to B5,
 respectively). This is a standard band selection for the source detection
 procedures in SAS,
 for which the  source significance (detection likelihood) calculation is well
 established.
 For the PN images in band B1 we 
selected only single-pixel events (PATTERN=0), 
while single
and double-pixel events (PATTERN$\le$4) were included in
 the other energy bands.
Such a PATTERN selection was motivated by significantly higher noise from
particle background
in the softest PN band.
 We  selected 1--4 pixel events
(PATTERN$\le$12) in all five   bands of the MOS images.
For each image we  produced the 
 exposure map corrected for vignetting, the background and sensitivity maps, and  the corresponding detector
 mask to
 define 
 source detection areas. 
 These maps were used as inputs for our source detection procedures
  in each SNR field.

\subsection{The X-ray source catalogs}
\label{catalogues}

To detect the X-ray sources in the observed fields,
we  run the standard SAS source detection procedures searching simultaneously
in a total of 15 images for each field
 (the 5 energy bands B1 to B5 in each of the 3 EPIC cameras).
 The SAS task {\tt emldetect} was used to determine the parameters
of  the detected sources (significance, count-rates, fluxes, and
 hardness ratios)
  by using a maximum likelihood fit.
 We  selected the 
 likelihood threshold value of mlmin=7, which 
allows detection of even very weak sources
(S/N$\approx$3$-$4; see the SAS manual).

To convert the vignetting-corrected count rate to flux for each source 
 in each
 energy band, we estimated 
the energy conversion factors for the PN and MOS medium
 filters 
 using W3PIMMS\footnote{http://heasarc.gsfc.nasa.gov/Tools/w3pimms.html} (Table~\ref{ecf}). 
 The cataloged values are calculated for a power-law model with 
a photon index of 1.7 absorbed by the    
$N_{\rm H}$ columns of 1.5 and $2 \times 10^{22}$\,cm$^{-2}$ for G27 and G28,
respectively\footnote{These absorption columns are
 equal to the Galactic
 H{\sc i}  columns measured in the directions of the two SNRs (Table~\ref{ecf}).}.
This model fits well the majority of the detected sources, i.e,
 intrinsically medium-hard sources, although we found that the inferred fluxes are not very
 sensitive to the spectral model selection. For example, for soft sources
 (e.g.,  best-fitted by a black-body model) our cataloged flux values are
 overestimated by 
 only $\sim$10\%. On the other hand, for significantly harder
 sources (for example, a power-law with a photon index of $\sim$1) the
 cataloged flux values are underestimated  by approximately 30\%.
  The calculated fluxes are actually much more sensitive to the assumed
 absorption  columns. For example,  
 nearby stars would be much less absorbed,
 but even if we assume an $N_{\rm H}$
 column of only  $5 \times 10^{20}$\,cm$^{-2}$, the flux values listed in
 Tables~\ref{cat1} and \ref{cat2} would be overestimated by approximately
  a factor of 2, which, however, does not affect the source classification significantly
 (Section~\ref{class}).

After compiling the list of X-ray sources detected above the
 $3\sigma-4\sigma$
threshold in both SNR fields, we compared the X-ray positions with optical and
 near-infrared (NIR)
data to determine possible offsets of the two {\sl XMM-Newton} pointings. 
We  cross-correlated the  X-ray positions with optical 
 \citep[USNO-B1;][]{2003AJ....125..984M} and NIR \citep[2MASS;][]{2006AJ....131.1163S}   catalogs using 
a distance of $< 3\times$ the combined  X-ray and optical/infrared positional error to
 associate the  
optical/infrared and the X-ray source.  
We used the  identified optical/infrared
 counterpart candidates 
 to calculate  
 mean offsets in RA and DEC, but did not notice any systematic shift (within the uncertainties) with respect to
 the optical/infrared data.  
A standard 1$\sigma$ (residual) systematic positional error 
 of $1\farcs5$ \citep{2003AN....324...89W} was then added quadratically 
 to the statistical positional uncertainties of the cataloged sources.

We detected 11 point X-ray sources in the field of G27.8+0.6, and 28 sources
in the G28.8+15 field. The results  are summarized in  Tables~\ref{cat1} to \ref{opt2}.
Tables~\ref{cat1} and \ref{cat2}  list the X-ray properties of these sources.
The source identification number is listed in column 1 (the same source
number is also used in Figure~\ref{radio-xmm1}),
 while the position 
  and the
   total positional error, which includes the statistical and
systematic errors, are listed in clumns 2 and 3, respectively.
The source detection likelihoods in all 15 images are added and transformed to the equivalent single-band detection likelihood, which is listed in column 4.
 Columns 5 and 6 list the   
total number of the accumulated counts and  corresponding error,
 and the total flux and error in the
 0.2--12.0 keV band. The fraction of the soft (0.2-4.5 keV) flux, used to
 calculate the ratio of the X-ray to optical flux (see Section~\ref{class}),
 is listed in column 7.  
 The 
hardness ratio, defined as $HR = (R_{\rm 2-12\,keV} - R_{\rm 0.2-2\,keV}) / (R_{\rm 2-12\,keV} +
R_{\rm 0.2-2\,keV}$)
(where $R$ is the count rate) and errors, all from the combined
 PN, MOS1, and MOS2 data, are listed in  column 8.
Visual inspection of the
  broad-band images revealed a marginaly extended source (src. \#1)
 in the field of 
G27.8+0.6. We discuss this source in Section~\ref{PWNcandidates}.

Tables~\ref{opt1} and \ref{opt2} 
list the results of the cross-correlation with  the USNO-B1,
 2MASS,
and radio \citep[NVSS,][]{1998AJ....115.1693C} catalogs, which are  
described in detail in 
Section~\ref{class}.
   The  detected USNO-B1/2MASS counterparts are listed in column 2,
 together with
   the  offsets from the X-ray position (3), and the corresponding X-ray to
 optical flux ratio (4).
 The tentative source classification is given in column 5,
and we discuss it further in Section~\ref{class}, while column 6 lists the probability of the optical/NIR association.

The point-source sensitivity limit of the {\it XMM-Newton} observation in the
field of 
G27.8+0.6 is about  $1.5\times 10^{-14}$ erg~cm$^{-2}$~s$^{-1}$ [corresponding to an unabsorbed luminosity of (1--3)$\times10^{31}$ erg s$^{-1}$ at the
 estimated SNR distance of 2--3 kpc and $N_{\rm H}= 1.5 \times 10^{22}$ cm$^{-2}$], and
slightly deeper across G28.8+1.5 [$(1-1.5)\times 10^{-14}$ ergs~cm$^{-2}$~s$^{-1}$ or $(2-6)\times10^{31}$ erg s$^{-1}$ at the
 distance of 3--5 kpc and  $N_{\rm H}= 2 \times 10^{22}$ cm$^{-2}$  estimated for this SNR]. The brightest source in the G27.8+0.6 field
($\sim2\times 10^{-12}$ ergs~cm$^{-2}$~s$^{-1}$) is an order of magnitude brighter than any of the
other detected sources, while in the second field 
the  measured absorbed fluxes range from $1 \times 10^{-14}$ to $3
\times 10^{-13}$ erg~cm$^{-2}$~s$^{-1}$.

Figure~\ref{radio-xmm1}   shows
  the radio images of the two SNRs, and
 the broad band (0.2--12.0 keV) images of both
 fields, combining PN, MOS1, and MOS2 data.
The detected X-ray sources are shown as circles in each field.
The  top left panel shows
 the VLA (1.4 GHz) image of G27.8+0.6. Its bright central  region is 
 clearly visible as well as a
larger  ($30'\times50'$)  faint extended component. The fainter extended
component is  also seen at
 lower
frequencies in the single-dish radio observations \citep{1984A&A...133L...4R}.
The bottom left panel shows the region
of G28.8+1.5 observed by VLA at 0.325 GHz. Both VLA images are extracted from 
 MAGPIS \citep{2005AJ....130..586W}. Sources 3 (in G27.8+0.6) and 6 (in G28.8+1.5) appear to be well centered within the bright radio emission regions (see also Section~\ref{PWNcandidates}).

Since the {\sl XMM-Newton} observations of the G27.8+0.6 and G28.8+1.5 are very short, the
 small number of 
accumulated counts limits  spectral analysis,
 even for the brightest sources. However, the hardness ratios suggest that
 we detected different source populations:
the sources  exhibiting intrinsically
hard spectra (e.g HR$> 0$),
and several extremely soft sources with most of the X-ray emission below
0.5 keV (e.g HR$\sim -1$).


\begin{figure}

\includegraphics[height=6.8cm,angle=0]{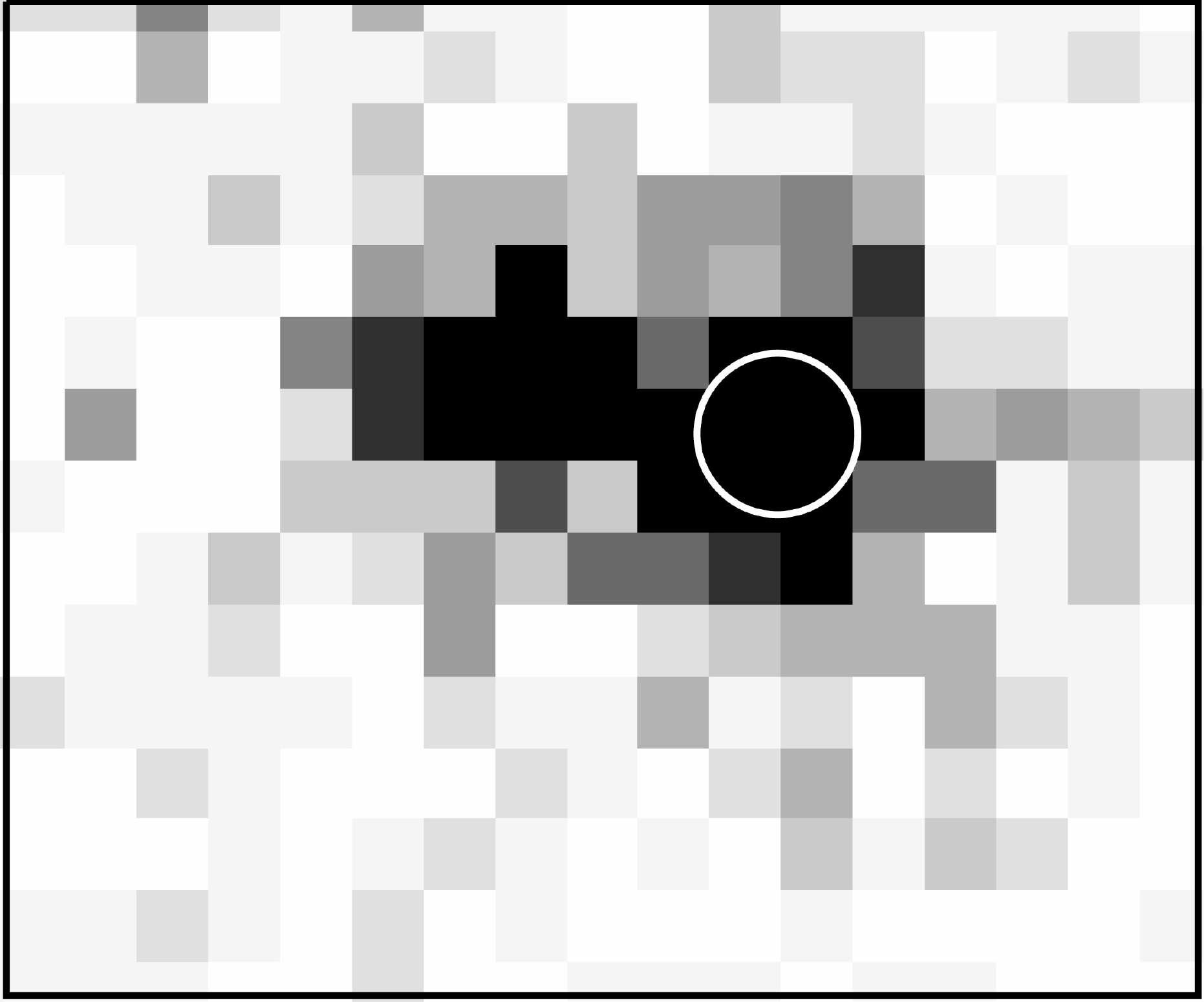}
\includegraphics[height=6.8cm,angle=0]{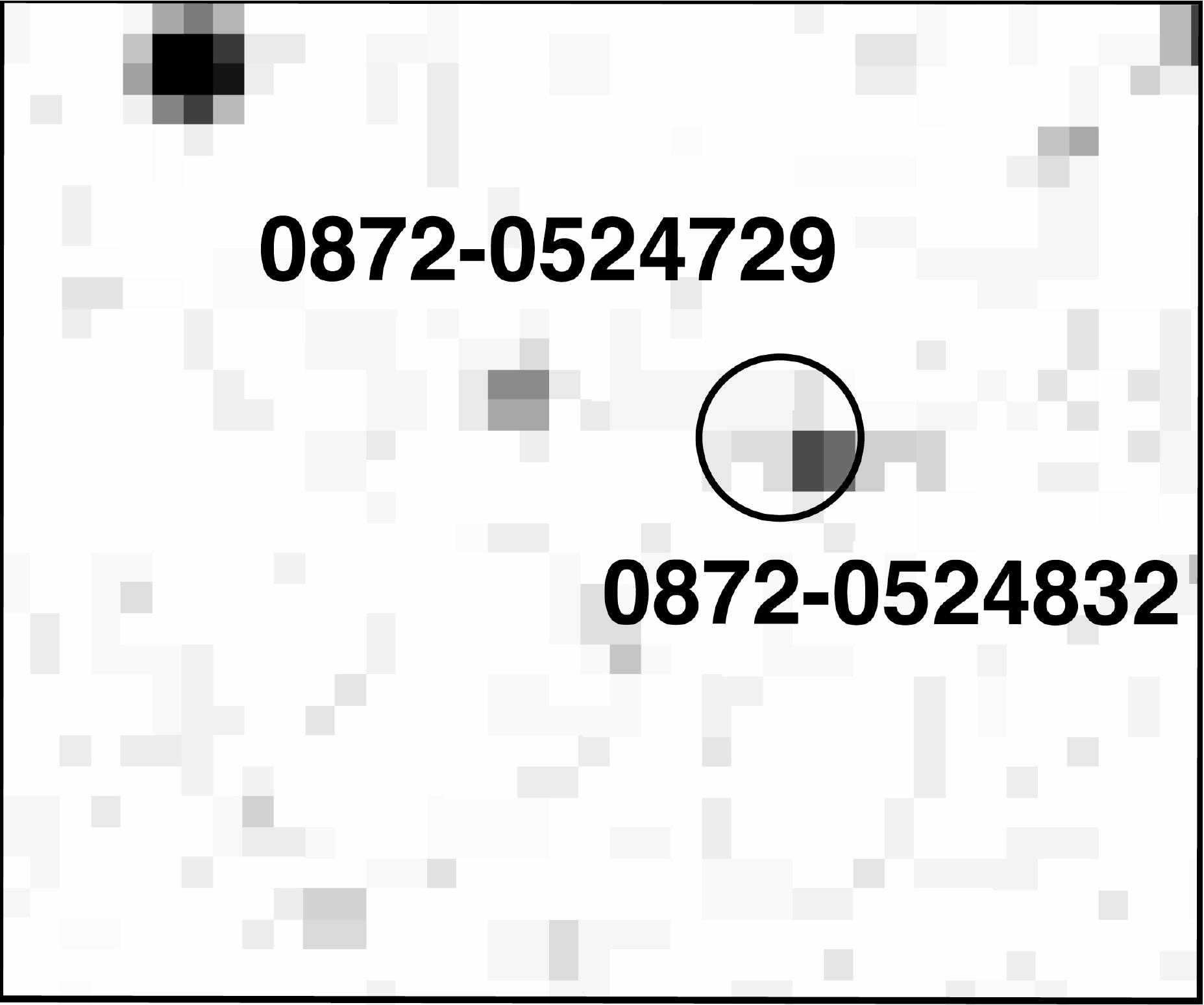}
\caption{The combined MOS1+MOS2 image (left)  and DSS image (right) showing  the region around the
 brightest point source (src. 1) in the field of G28.8+1.5. The black circle
 in the DSS image shows the position of the X-ray source as determined by our
 source detection procedure (R.A = 18:39:03.041, Decl. = $-$02:42:01.94).
 The radius of
 the circle is 4.5$''$ (3$\sigma$ positional error).
 The two USNO-B1 sources, which are most probably the optical counterparts
 of the
 detected X-ray emission, are also marked in the DSS image (see text for more
 detail). The 3$\sigma$ error circle of the cataloged X-ray position is also indicated on the MOS image
 (as the white circle). 
\label{optical-src1}}
\end{figure}


\subsection{Cross-correlation and classification of X-ray sources}
\label{class}

We cross-correlated all the X-ray sources listed in Tables \ref{cat1} and \ref{cat2}  with  the
 USNO-B1, 2MASS, and NVSS
catalogs.

We found  USNO-B1 counterpart candidates for each of the 28 X-ray sources
 detected in the
G28.0+1.5 field, and all except two sources (src. \#1 and \#7) in the field of
G27.8+0.6.
Although we find
 most of the candidate optical counterparts within the $\sim2\sigma$
 positional uncertainty of the X-ray source\footnote{All the optical candidate counterparts are
  within the $3\sigma$ of the X-ray positional error.}, there is a high probability of a chance coincidence, due to a broad PSFs of the {\sl
  XMM-Newton} detectors and a high density of stars in the Galactic plane.

Based on the optical source density ($\rho =0.0080$ and $\rho=0.0077$
 field sources
in a $1''\times1''$ box in the field of
G27.8+0.6 and G27.8+1.5, respectively), we calculate the
probabability\footnote{We estimate it as the probability of finding zero field sources in the circle of  radius $d$, $P={\rm exp}(-\rho \pi d^2)$, where $d$ is
 the distance between the optical and X-ray positions.} that
 each listed  optical source is the true counterpart (Tables~\ref{opt1}
 and \ref{opt2}).

To classify the sources we
 also used the  
X-ray-to-optical flux ratio calculated as $ {\rm log}({ f_{\rm x}/ f_{\rm opt}}) =  {\rm log}(
f_{\rm x}) + (m_{\rm B2} + m_{\rm R2})/(2\times2.5) + 5.37$, following
\citet{1988ApJ...326..680M}.  In this formula $f_{\rm x}$ is measured in the first four
bands (0.2--4.5 keV) because the source classification scheme of \citet{1988ApJ...326..680M} is based
on the relatively soft X-ray band (0.3--3.5 keV) of the {\sl Einstein
  Observatory}.
We used the average of the B2 and R2 USNO magnitudes instead of (mostly unavailable) V magnitudes, but the difference between these magnitudes for the main sequence stars should be less than $\pm0.1$ \citep{2000asqu.book.....C}. We have verified this for several stars in our catalogs, for which the USNO and the position of the optical source with the cataloged V magnitude are sufficiently close (i.e., the three magnitudes are most likely measured from the same source). However, the difference between  the averaged USNO and V magnitudes for sources classified as AGN/CVs seems to be slightly larger (up to 0.5) but still not large enough to change our source classifications.

The cross-correlation with the 2MASS catalog yielded 9 and 23 candidate 
 counterparts
 for the G27.8+0.6  and G28.8+1.5 field, respectively.
 We found only one radio
counterpart  -- the 3.6-mJy NVSS source
183932--023919 -- coinciding with the X-ray source \#27 from
 the G28.8+1.5 field.

We  used
the multi-wavelength data to classify the cataloged sources.
In particular, we used the X-ray hardness ratios and optical and NIR properties to identify
main-sequence stars and other source classes in our sample. 
Main-sequence stars are expected to have significantly smaller
values of log($f_{\rm x}/f_{\rm opt}$). Depending on the stellar type,
the log($f_{\rm x}/f_{\rm opt}$) for stars is in the range --5 to --0.5,
 while galaxies, AGNs,
 X-ray
 binaries
(XRB) and cataclysmic variables (CVs) are expected to have significantly stronger X-ray
emission comparing to optical \citep[e.g., $-1<{\rm log}(f_{\rm x}/f_{\rm opt})<1$ for AGNs;][]{1988ApJ...326..680M}.
We also used
NIR
colors, in addition to the optical data, to confirm the stellar classifications
\citep[e.g., see Table 6 in][]{2000AJ....120.2615F}. 
We used these selection criteria and tentatively classified the majority of the detected sources in both SNR fields.

\subsubsection{Sources in the G27.8+0.6 field}

 In the field of SNR G27.8+0.6 we  classified 8 sources as main sequence stars (Table \ref{opt1})
based on their 
 log($f_{\rm x}/f_{\rm opt})$ and relatively soft X-ray spectra.
  The
NIR colors support the
 proposed spectral classification,
suggesting that the majority of these objects are nearby stars.

There is an optical counterpart candidate within the $3\sigma$ error circle of source \#3.
 Although the X-ray-to-optical ratio suggests a
stellar classification, the distance between the optical and X-ray position is
almost 6$''$, making the chance coincidence very likely (see Table~\ref{opt1}). In addition, the 
X-ray spectrum of this source is significantly harder than the typical stellar
spectra. We identified a 2MASS source  closer to
the X-ray position (off by $\sim$2$''$), which suggests that the source is either an AGN, CV or a
faint X-ray binary.

  We note that the probability of a chance coincidence between
the source \#8 and its possible optical counterpart is also very high (about 40\%). However,
 this
source, which exhibits a soft X-ray spectrum, is detected at a large offset
from the center of the field of view, which reduces the accuracy of the X-ray position.

 Two sources (src. \#1 and \#7), which display relatively hard X-ray
 spectra in comparison with identified field stars,  do not have any cataloged optical or infrared counterparts. 
While one of these objects (src. \#7) is detected as a weak X-ray source ($S/N
\sim 4$, see Table~\ref{cat1}),
  source \#1
is the brightest source in the field.
 Furthermore, the morphology and spectral properties of this source 
 suggest that this
object could be a NS created in the explosion of G27.8+0.6 and moving
 away
from its birth site. We discuss this source in more detail in
 Section~\ref{PWNcandidates}.

\subsubsection{Sources in the G28.8+1.5 field}

 All of the 28 sources detected in the G28.8+1.5 field
have  optical candidate counterparts within the $3\sigma$ X-ray
 error circle. Based on the X-ray, optical and NIR data, we
classified 11 sources as stars, while 3 objects are stars or galaxies,
 according to their NIR colors. 
The remaining 14 sources have significantly larger X-ray-to-optical flux
 ratios and
harder X-ray spectra. The surface number density of these hard sources is
approximately the same as that of the background extragalactic sources
\citep{2005ApJ...635..214E}, suggesting that most of them are AGNs, although
 some of them
might be either faint  X-ray binaries, or, more likely, cataclysmic
variables, which are considered to be prime candidates for faint Galactic hard
X-ray sources \citep[e.g.,][and references therein]{2005ApJ...635..214E}.

With approximately 400 accumulated counts
in the PN data, source \#1 (Table~\ref{cat2})  is the only source in the field of G28.8+1.5
 sufficiently bright for spectral analysis. We found the best fit model for this source to be an absorbed MEKAL ($kT=0.7-0.9$ keV) plus a hard component
 represented
by a power-law model ($\Gamma = 2.2 - 2.3$). The low value of the absorption
column (below $8\times10^{20}$ cm$^{-2}$) indicates that the source is much
closer than the SNR. We found that the flux obtained from the spectral
 fit, $f_{\rm X}=(0.6-1.3)
\times 10^{-14}$ ergs cm$^{-2}$ s$^{-1}$ (0.2$-$12 keV), is slightly lower than
 the cataloged value (which assumes a PL model with a
significantly larger absorption column).
We also obtained an acceptable
 fit for this source using a two-temperature plasma model ($kT\sim 0.1$ and
$kT\sim 0.8$
 keV), typical of active late-type stars
 \citep[e.g.,][]{2004A&ARv..12...71G},
 but could
 not find an acceptable fit for a single-component or any combination of the
 power-law or blackbody models.  
However,
as  Figure~\ref{optical-src1} indicates, there are probably two X-ray sources (most likely stars) at this position, which are difficult to resolve with {\sl XMM-Newton}.

\section{PWN/NS candidates}
\label{PWNcandidates}

\subsection{A PWN candidate in G27.8+0.6}
\label{SNR-g27}

\begin{figure}
\begin{center}
\includegraphics[height=4.6cm,angle=0]{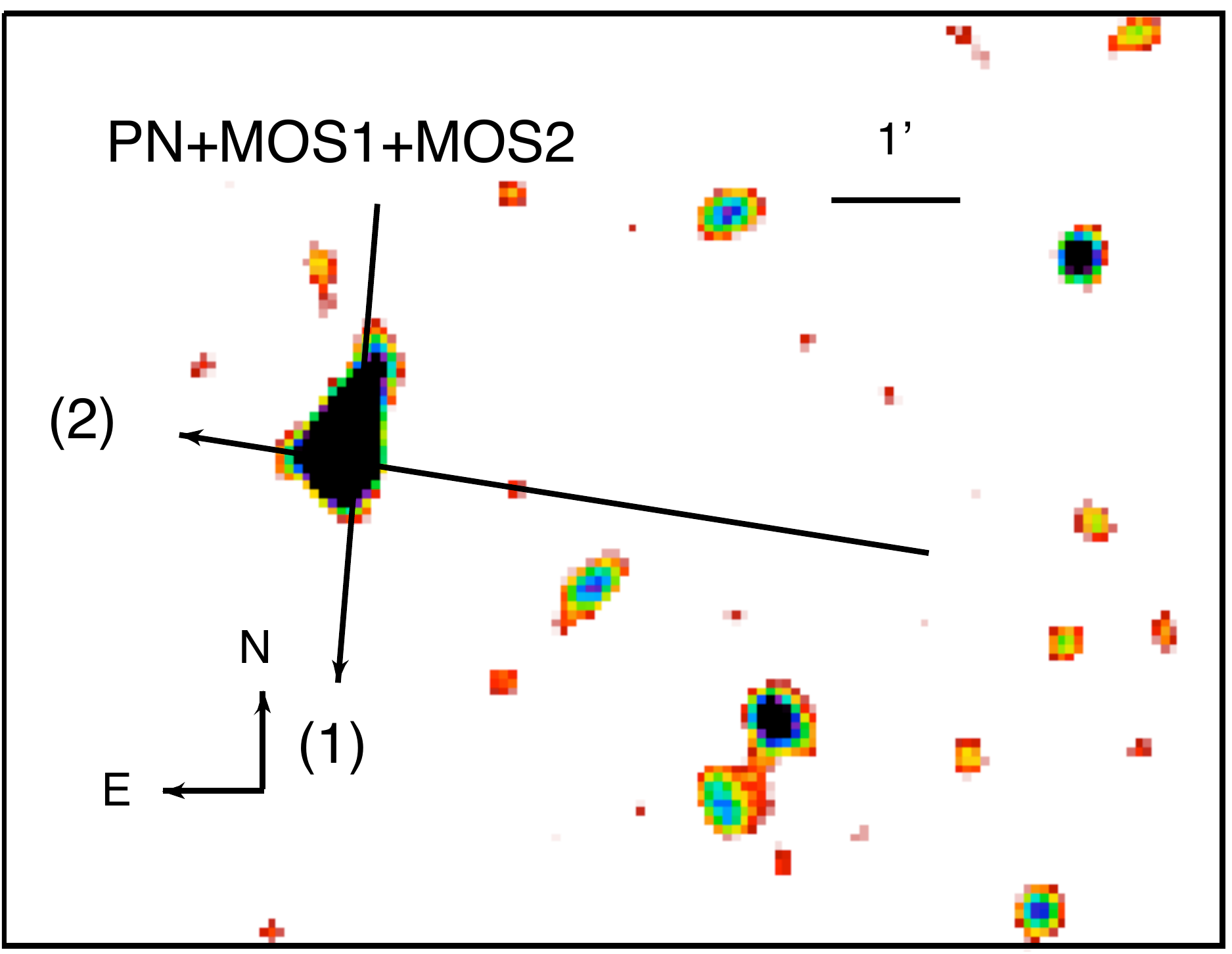}
\includegraphics[height=4.6cm,angle=0]{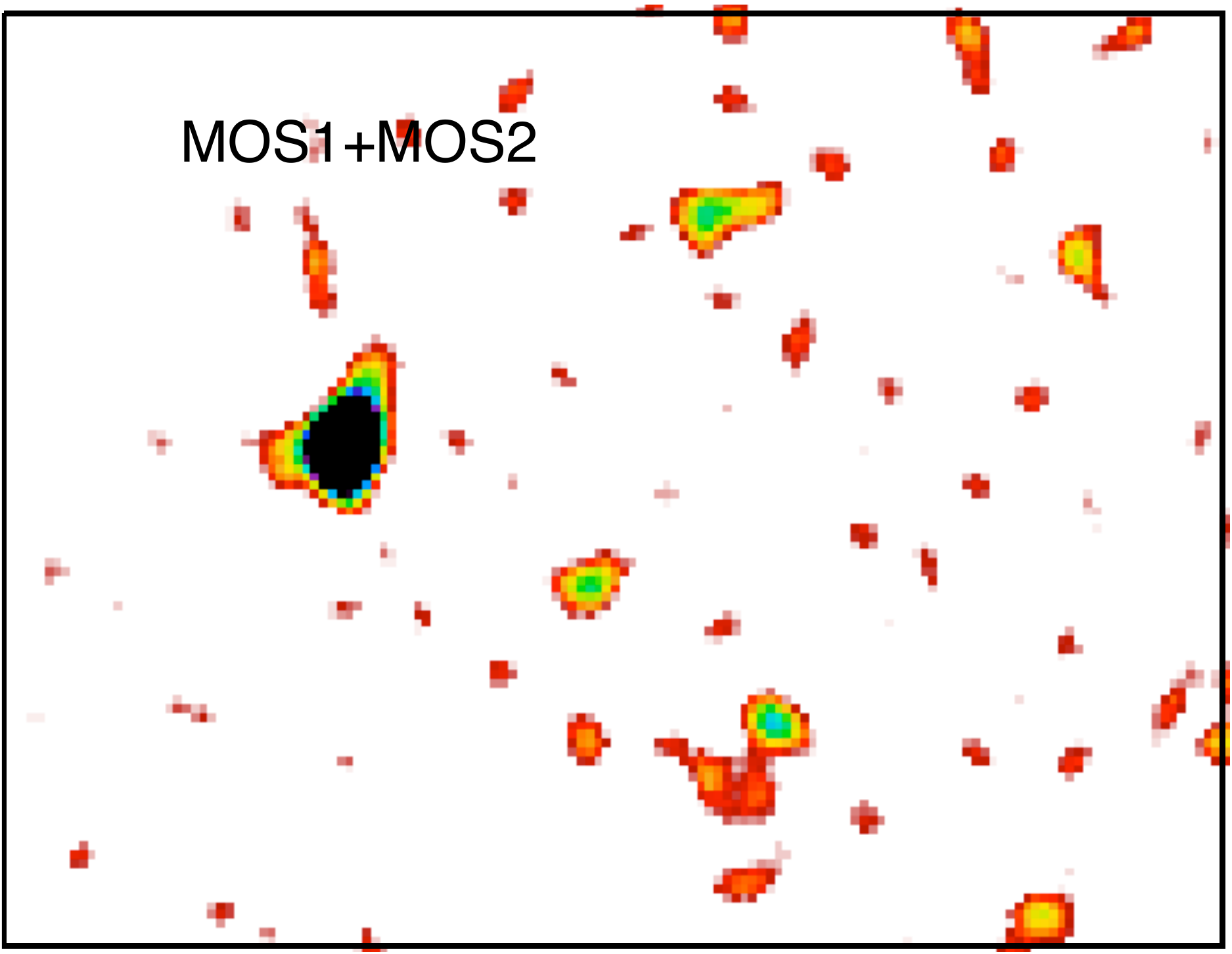}
\includegraphics[height=4.6cm,angle=0]{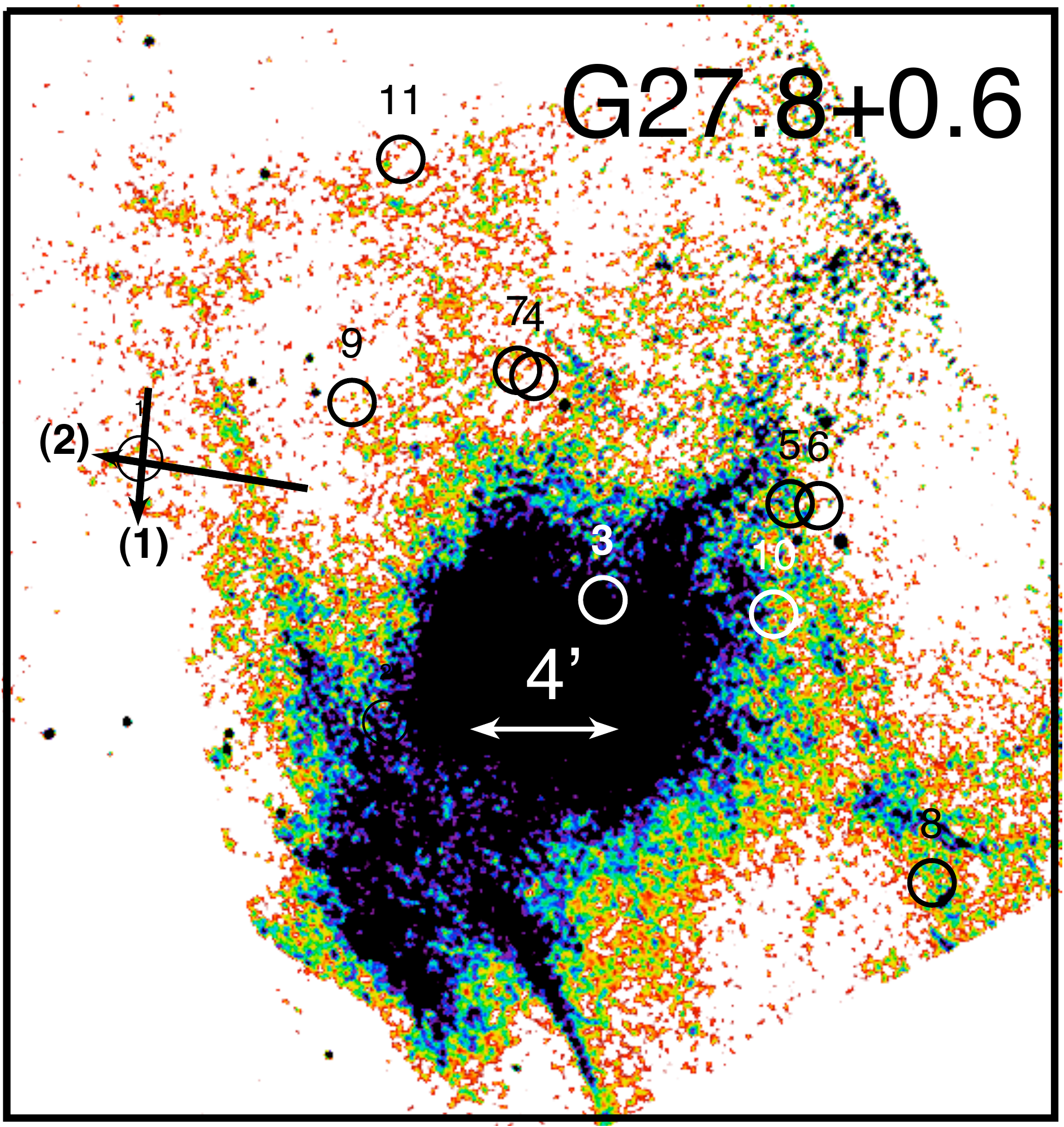}
\end{center}
\caption{
 Broad-band (0.7--10 keV) EPIC image of
 the PWN
 candidate (Src.1 in Tables~\ref{cat1} and \ref{opt1}, see also Figure~\ref{radio-xmm1})
 combining the PN, MOS1, and MOS2 data (left) and only MOS1 and MOS2 data (middle).  The images are smoothed with a Gaussian of FWHM 16$''$.
 The pixel size is 4$''$.
 The intensity scale is linear from 0 to 1.42838
 cts~s$^{-1}$~pixel$^{-1}$. 
  The two
 arrows marked (1) and (2) show two possible directions of motion of the
 pulsar
candidate (see text for detail). 
 A VLA image of G27.8+0.6 at 1.4 GHz (right) is
extracted from
  MAGPIS \citep{2005AJ....130..586W}. The sources detected in the {\sl
    XMM-Newton} observations are overlaid on the radio map. The position of
  the PWN candidate (Src. 1) is marked by the same two arrows shown in the
  EPIC image. 
 \label{g27-pwn}}
\end{figure}

 No extended diffuse X-ray emission coming from the SNR shell or its central
 regions is visible in the EPIC images of G27.8+0.6. 
This is not surprising because 
 after filtering the data affected by flaring the remaining
good-time exposure was only about 4.9 ks for PN, 10 ks for MOS1, and 8.4 ks for
 MOS2
  (Table~\ref{obslog}), making it very difficult to detect any faint,
large-scale  emission above the background.

\begin{figure}
\includegraphics[height=5.6cm,angle=0]{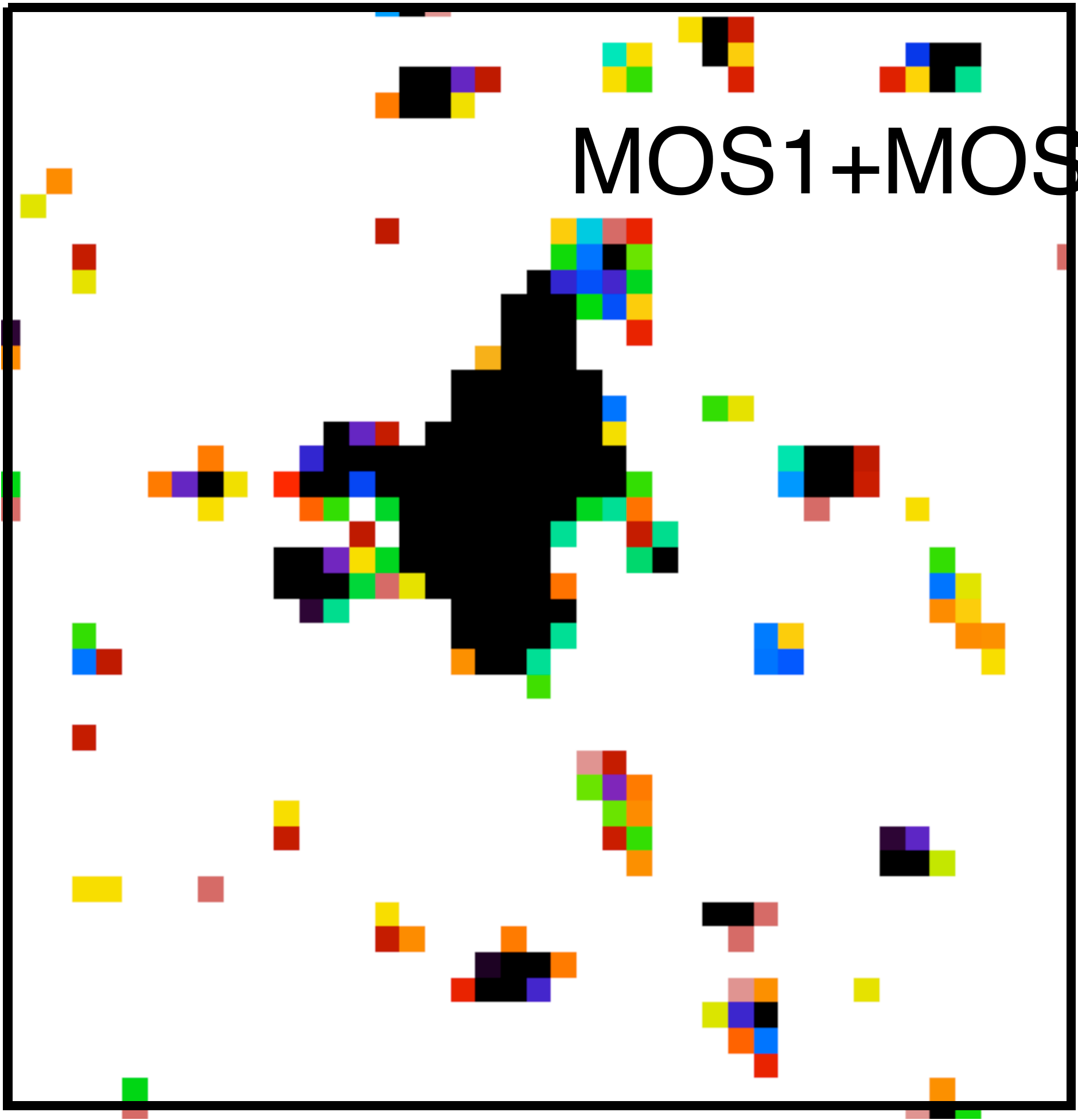}
\includegraphics[height=5.6cm,angle=0]{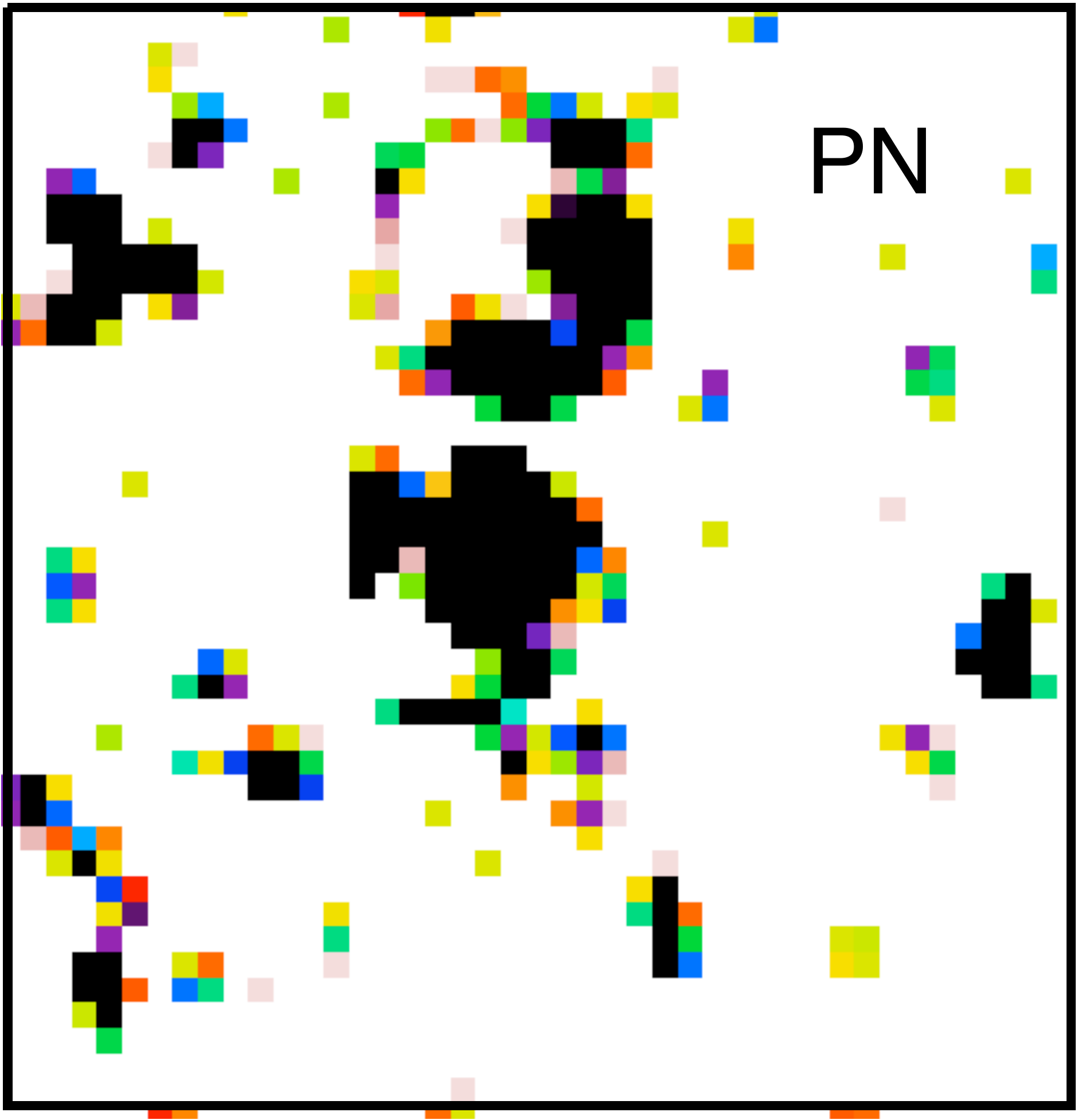}
\includegraphics[height=5.6cm,angle=0]{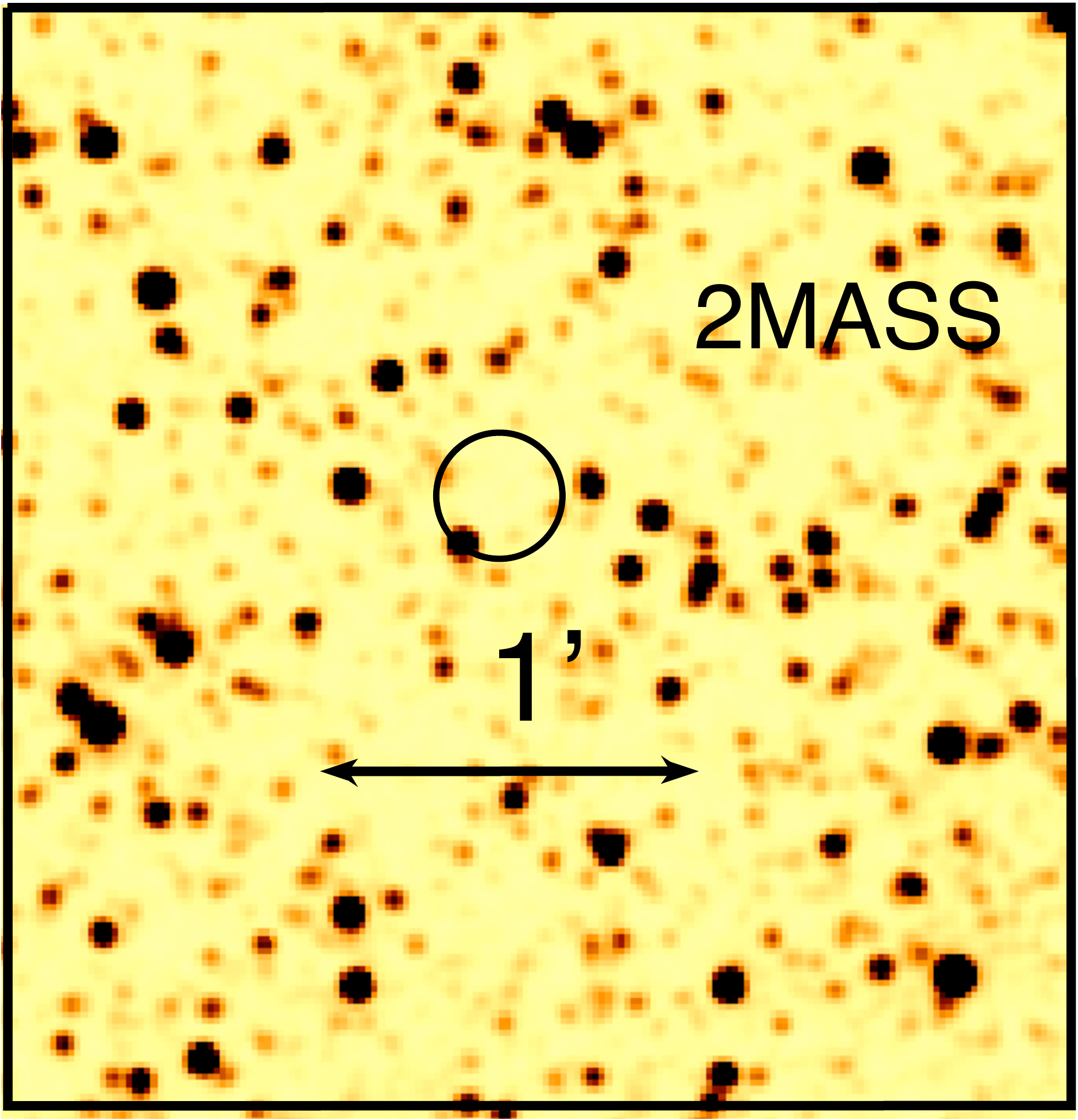}
\vspace{-0.5cm}
\caption{Broad band (0.7-10 keV), binned 
 MOS1+MOS2 (left) and PN (middle)
 images 
showing the region around the detected PWN candidate, which consists of a bright point source and possible extended emission.
After screening for the periods of high background, the exposure times were
4.9 ks for PN, 10 ks for MOS1, and 8.3 ks for MOS2. 
 The NIR J-band (2MASS) image of the same region (right)
 shows no bright sources that could be identified with the X-ray emission (see text). The circle with a radius of 10$''$ marks the position of the source.  
\label{g27-ind}}
\end{figure}

However, we have detected a bright, possibly extended source 
 offset by  $\sim$20$'$ (or $\sim$12 pc at
 the  SNR's distance of 2--3 kpc)   from
 the presumed SNR's center (Figure~\ref{g27-pwn}).
Although the source was not flagged as extended in our source detection procedures, the smoothed (Figure~\ref{g27-pwn}) and binned (Figure~\ref{g27-ind}) images
 suggest that there is
 a small
  elongation in the North direction 
  and perhaps some weak
  emission  to the East from the point source. 
Although the  number of  accumulated counts is small (see Table~\ref{cat1})
 because of
the short  exposures, the
  extended structures are clearly visible, not only in the combined data but also in individual PN and MOS images.
We note that there is a CCD gap and a bad column across the source in the
 PN data,
 which cuts off
a significant fraction of the point source, and perhaps also a part of the
candidate nebula\footnote{We have excluded all border pixels from these images (by selecting FLAG=0), to make sure that no artificial structures are produced in the vicinity of the source.}. Since the source is detected 10$'$  off-axis, 
 the EPIC PSFs are  elongated in the North-South direction.
However, the entire source extent is not likely to be due
to the PSF distortion\footnote{For example, compare with the PSF of source 2 in the G28.6+1.5 field, which, at approximately the same off-axis angle, is less elongated. }. 
In addition to possible extended emission in the vicinity of the point source, there are several relatively bright blobs visible in the Southwest and Northwest  directions, forming a triangular-like shape with the point source at the vertex. These blobs (with a signal to noise of $\sim$3--4, estimated in an aperture of $40''\times40''$ in the 0.7--10 keV, measuring the background from the nearby area) might be part of extended emission at a distance of $\sim5'-10'$ from the source.  Figure~\ref{g27-ind}  (right)  shows a NIR (2MASS)
 image of the
 same region with no optical counterpart at the position of the X-ray
 source. We have searched optical catalogs and found no sources within 15$''$ radius with $I<20^{\rm mag}$ \citep[and down to the limiting magnitude of the USNO-B1 catalog, $M_{\rm V} \sim 21$;][]{2003AJ....125..984M}.

To determine the spectral properties of our PSR candidate,
we extracted the source spectra from the PN and MOS data
and group them with a minimum of 10 counts per bin.
However, the small number of the accumulated photons ($\sim$60 in PN and
slightly more in MOS1+MOS2 after subtracting the background)
 in our short-exposure
observations, and a large ($\sim$50\%) background contribution, did not allow
us to  constrain well the fitting parameters. 
 
We have obtained an acceptable fit ($\chi^2/{\rm d.o.f} = 0.84/16$)
for an absorbed power-law
model  with the photon index $\Gamma=1.2\pm0.8$ and the hydrogen column density
 $N_{\rm H}= (3\pm 2) \times10^{22}$ cm$^{-2}$, consistent with the H{\sc i}
 value
   from \citet{1990ARA&A..28..215D}.
Using this model, we estimated the source flux to be about\footnote{this a factor of 2 below the flux listed in Table~\ref{cat1} due to using different model parameters}  $4\times10^{-13}$erg~s$^{-1}$cm$^{-2}$
 in the 1.5--10 keV band,
implying a luminosity range of $L_{\rm X}=(2-5)\times10^{32}$ erg s$^{-1}$
 for an assumed 
  distance of 2--3 kpc.

\begin{figure}[h]
\includegraphics[height=7.0cm,angle=0]{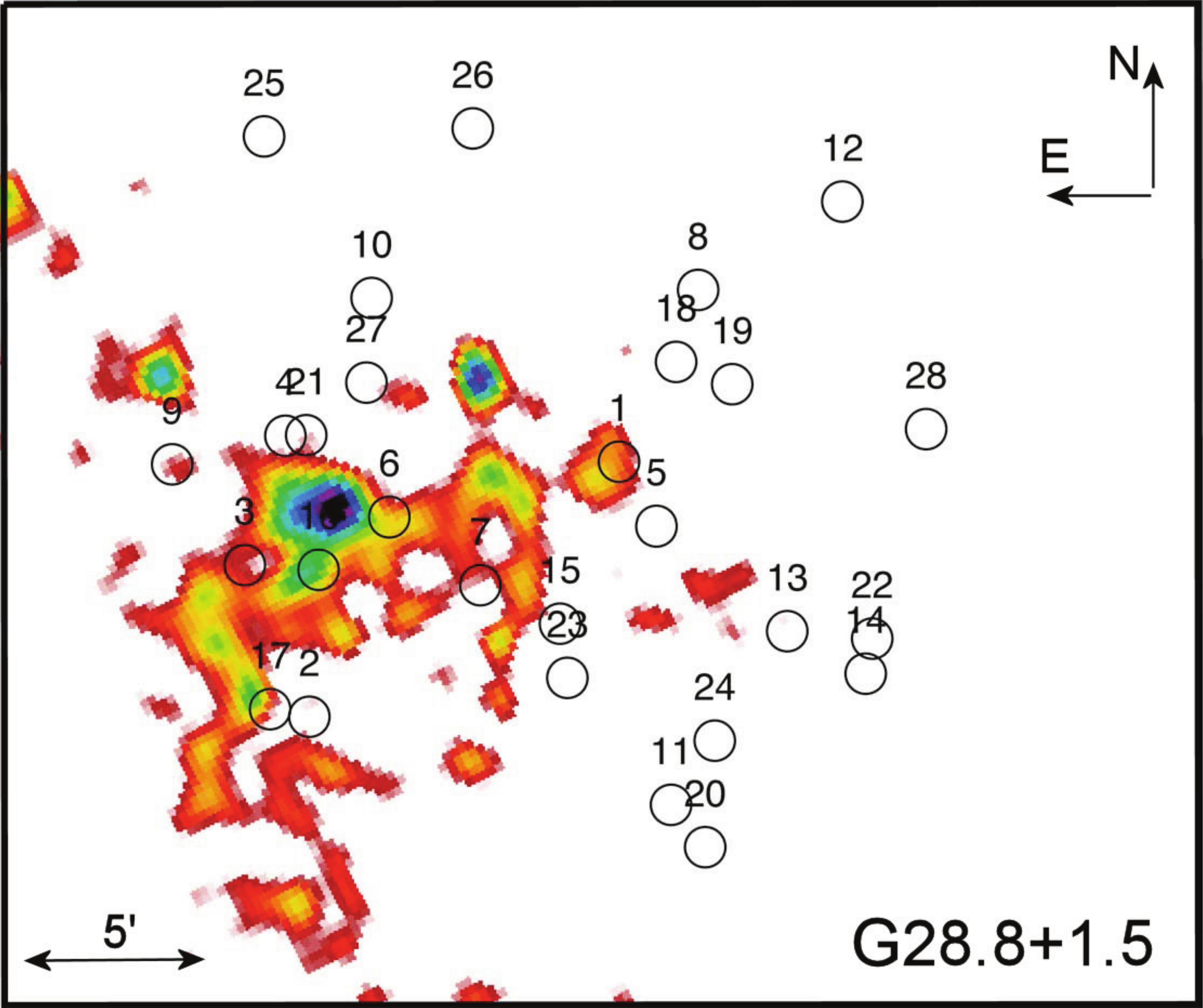}
\includegraphics[height=7.0cm,angle=0]{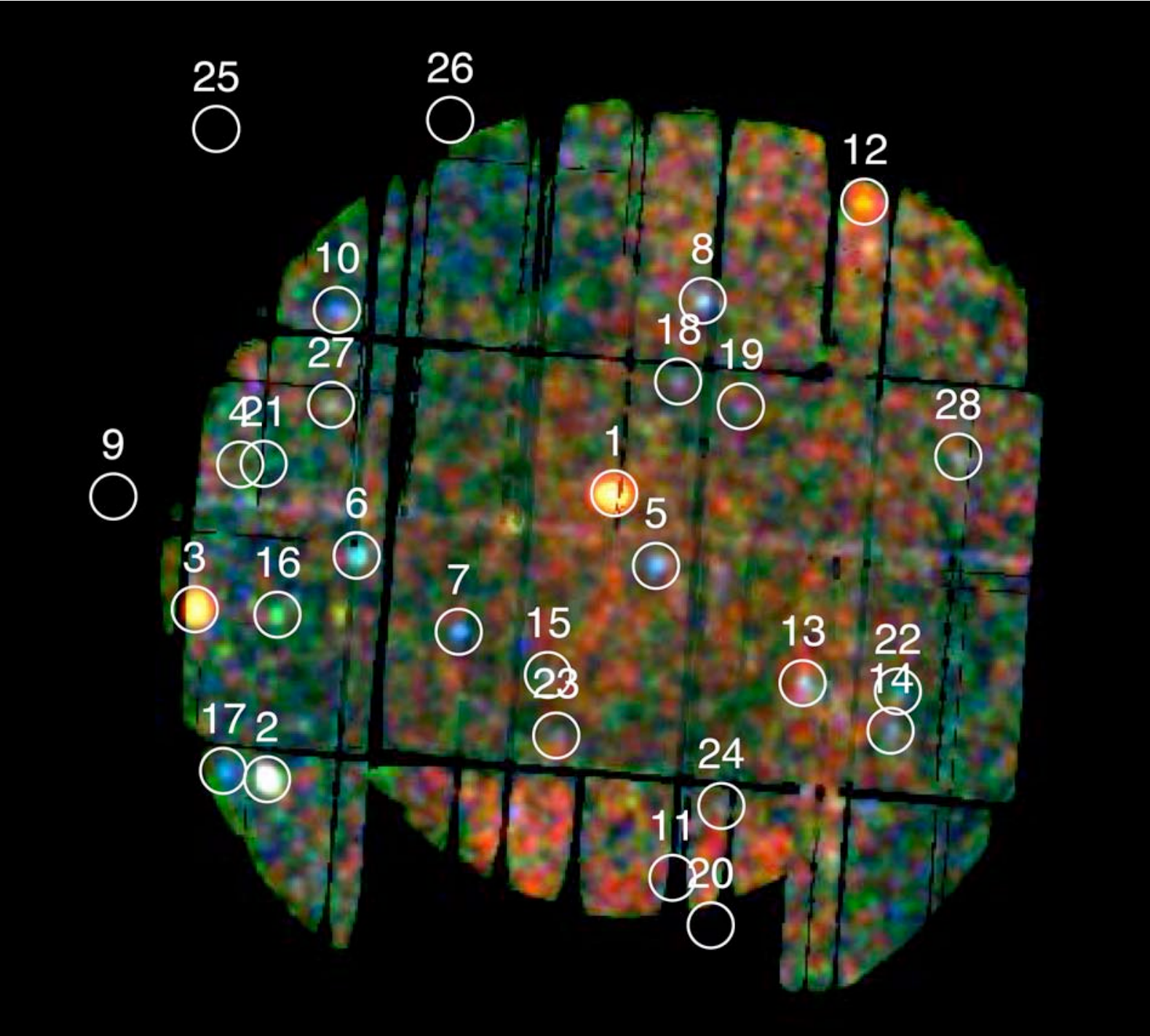}
\caption{{\bf Left:}
 VLA image of  
  G28.8+1.5 at 0.325 GHz
extracted from
  MAGPIS \citep{2005AJ....130..586W} probably showing part of the SNR shell.
{\bf Right:}  Color-coded (red: 0.5$-$1 keV; green:
1$-$2 keV; blue: 2$-$4.5 keV), exposure-corrected and smoothed EPIC image  of the same region showing 
soft diffuse emission across the SNR and around its center.
 The cataloged X-ray sources are overlaid on both images.  
 \label{diffuse} }
\end{figure}

The right panel of Figure~\ref{g27-pwn} shows
 the radio map of G27.8+0.6 and the
position of the PWN candidate with respect to the SNR center. The two arrows
(also shown in the EPIC image on the left) indicate two possible velocity directions of
the pulsar candidate, based on the PWN shape.
 If the blobs extending westwards are part of a ``tail'' formed by a
bow-shock, due to the pulsar's supersonic velocity, then, according to the
existing MHD  models \citep[e.g.,][]{2005A&A...434..189B}, the pulsar travels 
in the direction of the arrow marked 2, i.e., away from the
center of the brightest part of the extended radio emission,
 which probably coincides with the site of the SN
 explosion. The extended emission to the North could be then
 explained as equatorial emission \citep[a torus;][]{2000ApJ...536L..81W}.   
However, we cannot exclude the possibility that the pulsar candidate is
 moving in the 
direction of arrow 1 (i.e., the ``tail'' extends to the North, and the blobs are not associated with the PWN candidate).   Although, in this case, the pulsar's velocity
 would be nearly perpendicular to the
 expected radial direction,  similar behavior was also observed  in other
 SNR-PWN systems and explained by 
a high initial velocity of the NS progenitor, which caused the SN explosion offset from the center of the wind-blown cavity and the  remnant's geometric center
 \citep[e.g., similar to the PWN in the IC443;][]{2006ApJ...648.1037G}.

We conclude that the lack of the optical counterpart, the X-ray spectral properties and possible extended emission surrounding the source
 are all 
consistent with the proposed PWN identification. If the nebula was formed by
the central pulsar associated with G27.8+0.6,
its velocity\footnote{The velocity is estimated from 
the remnant's age and the traveled distance of $\sim$12 pc if the pulsar's
birth place is at the SNR center.} of
around 100--300 km~s$^{-1}$  would be consistent
 with measured
 velocities
of other pulsars. 
  In addition, its position close to the host SNR
edge, suggests that perhaps the pulsar's velocity is at the transition from
being  subsonic in
the SNR interior to supersonic outside the SNR. 
 According to the current  models \citep[][and references
therein]{2005AdSpR..35.1123V},
 the pulsar's velocity becomes supersonic with respect to the ambient
 gas when the pulsar approaches the regions of cooler material near the  edge
 of the shocked ejecta (for a SNR in the Sedov phase this occurs at the $\sim
 2/3$ of the distance from the SNR's center).
 In this evolutionary stage, there are two 
PWN components: in addition to the X-ray bright PWN enveloping the moving
pulsar, there is also a relic PWN (center-filled radio plerion) formed in
the supersonic SNR expansion stage and still emitting synchrotron radiation at radio frequencies, with significantly longer lifetimes.
Although our X-ray and radio data seem to be entirely consistent with this picture, further observations are needed to confirm that the source is extended and to study its X-ray morphology in detail.

Finally, we note that we cannot entirely rule out the possibility that source \#3 is a NS associated with G27.8+0.6.  This source is located close to the center of the plerionic radio emission and its candidate optical/NIR counterparts are $\sim$6$''$/2$''$ away (see Section~\ref{class}). However, even in this case, source \#1 would remain a PWN candidate without a radio SNR association, which is a rather common occurrence \citep[e.g., see][]{2008AIPC..983..171K}.

\subsection{Supernova remnant G28.8+1.5}
\label{G28.8+1.5}

The bottom right panel of Figure~\ref{radio-xmm1} shows an amorphous region ($\sim$10$'$ in diameter) of extended X-ray emission around the central part of SNR G28.8+1.5.
 Although  faint, the diffuse emission has apparently a soft spectrum (suggesting a thermal plasma) as it can be seen   
 in the exposure-corrected\footnote{To produce the exposure-corrected
   images, we used the exposure maps corrected for vignetting.}
 color-coded EPIC image 
 (Figure~\ref{diffuse}, right).
   The comparison with
 the radio map (Figure~\ref{diffuse}, left) 
 shows that the X-ray emission comes from the
 region inside the bright
 rim of radio emission (i.e., perhaps part of the radio shell). 
In addition to shell-type, center-filled and composite
 (a combination of shell and
center-filled morphology) SNRs, there is around 8\% of SNRs with radio
shells filled with thermal X-ray emission,  called
mixed-morphology SNRs \citep{1998ApJ...503L.167R}. Although the  mechanism of
the central thermal X-ray emission is not well understood, almost all of the
mixed-morphology SNRs appear to be interacting with clouds
\citep[e.g.,][]{2001ApJ...552..175K}. While the {\sl XMM-Newton} data suggest
 that
G28.8+1.5 might be
 a mixed-morphology SNR, high spatial resolution observations are needed to
see whether  this emission can be resolved into possible
 point sources and truly diffuse 
emission coming from the SNR, and measure its spectrum.

There are two point X-ray sources close to the center of the diffuse
 emission,
source \#1, which we identified as a star (or, more likely,  a pair of
unresolved stars), and source \#5,
whose optical and infrared counterparts suggest an AGN
classification. However, 
 due to the broad PSF of the EPIC cameras,
 a {\sl Chandra}
observation would be needed to confirm some of these identifications.
 In particular,
we note that several sources (most of which are identified as AGNs) have large
offsets from their candidate optical counterparts (e.g., sources \#6, 8, 9,  17,
 18, 21, 22;
see Table~\ref{opt2}), which put their identifications into question.
Among these sources, particular attention deserves source \#6, which coincides with the brightest radio emission region that could be a radio PWN.

\section{Conclusions}
\label{conclusions}

We analyzed {\it XMM-Newton} observations of two center-filled SNRs,
 G27.8+0.6
and G28.8+1.5, in a search for their central pulsars/PWNe
and/or other types of NSs associated with these SNRs. 
We compiled the X-ray source catalogs, measured the fluxes and
hardness ratios, and also searched for cataloged  optical, infrared and radio
 counterparts,
 which we used to classify these objects.

We found that most  of the detected
sources in both fields are  foreground stars.
  The proposed classification of these sources is  based on hardness of their 
 X-ray spectra and optical and infrared properties of their candidate counterparts in the USNO-B1 and 2MASS catalogs.
The remaining sources are classified as AGNs (perhaps also including
 several galaxies and CVs),
although the separations between the X-ray and 
 optical/infrared positions for
these objects are significantly larger. Hence, some  of the proposed optical
 counterparts may be
 coincidental (e.g., some of these sources might be NSs).

We identified one possibly extended source (Src. \#1 in G27.8+0.6)
 as a PWN candidate associated with G27.8+0.6. This bright source does not
have an optical counterpart  and its morphology
 and X-ray
spectrum seem to be consistent with the proposed PWN classification. 
The position of the PWN candidate with respect to its host SNR
suggests a possible scenario:
the  alleged pulsar is moving away from the SNR center 
 with a velocity of 200$-$300 km~s$^{-1}$, and its position
 is near the edge of the
 shocked
 ejecta,
  approaching the cooler regions of the ambient material.  At this stage the pulsar's velocity becomes supersonic,
  and  a bow-shock nebula is being formed.
  A high-resolution timing
  observation  is
  needed to search for pulsations from this source.

Source \#3 in G27.8+0.6 and source \#6 in G28.8+1.5 could also be NSs associated with the SNRs, based on their locations within the  regions of brightest radio emission (radio plerions) combined with relatively low probability of optical/NIR associations.

We also detected faint extended emission around the central region of G28.8+1.5 (apparently inside its partial radio shell). If this emission is associated with  G28.8+1.5, it would put this source in a rare class of mixed-morphology SNR.

\acknowledgments{ This work was partially supported by NASA grants NNG05GR01G and NNX09AC84G.\,\,\,
 The publication makes use of data products from the Two Micron All Sky Survey, which is a joint project of the University of Massachusetts and the Infrared Processing and Analysis Center/California Institute of Technology, funded by the National Aeronautics and Space Administration and the National Science Foundation, and the USNOFS
Image and Catalogue Archive operated by the United States Naval
Observatory, Flagstaff Station\\ (http://www.nofs.navy.mil/data/
fchpix/). 
}

\clearpage

\begin{table*}
\scriptsize
\begin{center}
\caption[]{The {\sl XMM-Newton} observation log}
\vspace{0.4cm}
\begin{tabular}{cccccccc}
\hline\noalign{\smallskip}
\hline\noalign{\smallskip}
\multicolumn{1}{c}{SNR} & \multicolumn{1}{c}{Obs. ID.} &\multicolumn{1}{c}{Date} &
\multicolumn{2}{c}{Coordinates} & \multicolumn{1}{c}{EPIC PN} & 
\multicolumn{1}{c}{EPIC MOS1} & \multicolumn{1}{c}{EPIC MOS2}  \\ 
\noalign{\smallskip}
  & & & \multicolumn{2}{c}{RA/DEC (J2000)} 
& \multicolumn{1}{c}{T$^{a}$ (T$^{b}$)}
& \multicolumn{1}{c}{T$^{a}$ (T$^{b}$)}
& \multicolumn{1}{c}{T$^{a}$ (T$^{b}$)}\\
\noalign{\smallskip}
\noalign{\smallskip}\hline\noalign{\smallskip}
G27.8+0.6 & 0301880401 & 2005-10-04 & 18:39:49.9 & -04:24:00 &  4.93 (8.53) &
10.16 (10.16) &  8.36 (10.16) \\
G28.8+1.5 & 0301880901 & 2005-10-04 & 18:39:04.9 & -02:43:44 &  14.03 (14.93)
&   16.38 (16.56) &   16.56 (16.56) \\
\noalign{\smallskip}
\hline
\noalign{\smallskip}
\end{tabular}
\label{obslog}
\end{center}
$^{ a~}$ exposure time in units of ks after background screening\\
$^{ b~}$ total exposure time in units of ks  \\
\normalsize
\end{table*}

\clearpage

\begin{table}[t]
\caption[]{Count rate to  energy conversion factors (ECF;
 absorbed flux=count-rate $\times$ ECF) for the medium
 filter of the
 EPIC instruments in the energy bands B1--B5, assuming
            a power-law model with the photon index of 1.7 and Galactic
  absorption of $(1.5-2) \times 10^{22}$\,cm$^{-2}$ measured in the
  direction of G27.8+0.6 and G28.8+1.5 \citep{1990ARA&A..28..215D}.}
\vspace{0.5cm}
\begin{tabular}{lrrrrr}
\hline\noalign{\smallskip}
\hline\noalign{\smallskip}
\multicolumn{1}{l}{Detector} &\multicolumn{1}{c}{B1} &
\multicolumn{1}{c}{B2} & \multicolumn{1}{c}{B3} & 
\multicolumn{1}{c}{B4} & \multicolumn{1}{c}{B5}  \\ 
\noalign{\smallskip}
 &  \multicolumn{5}{c}{($10^{-12}$ erg cm$^{-2}$ ct$^{-1}$)} \\
\noalign{\smallskip}\hline\noalign{\smallskip}
 PN &  1.56 & 2.07 & 3.36 & 8.76 & 28.31 \\
        
 MOS &  6.22 & 6.44 & 8.19 & 19.88 & 94.95 \\
\hline
\end{tabular}
\label{ecf}
\end{table}

\clearpage

\rotate{
\begin{table}
\hspace{-3cm}
\scriptsize
\caption[]{X-ray sources detected in the field of SNR G27.8+0.6}
\begin{tabular}{cccccccccccc}
\hline\hline\\
  Src  & Position  & Err   & LoD & Cts & Flux & Soft fr. & HR \\
\hline \\
  & J2000   & arcsec   &  &  & erg s$^{-1}$cm$^{-2}$ &    
   &  &  &  & &  \\
\hline \\
1&18:40:37.49, -04:17:36.7 &1.7 &482.8 &340 $\pm$ 20 &2.07 $\pm$ 0.17E-12 &0.16 & 0.77 $\pm$ 0.16 \\
2&18:40:10.97, -04:24:42.8 &2.2 &37.3 & 50 $\pm$  9 &4.78 $\pm$ 2.99E-14 &0.70 & -1.00 $\pm$ 0.16 \\  
3&18:39:47.50, -04:21:25.0 &2.1 &26.2 & 57 $\pm$ 10 &9.96 $\pm$ 2.18E-14 &0.20& 0.98 $\pm$ 0.19 \\
4&18:39:54.95, -04:15:24.0 &2.5 &17.9 & 40 $\pm$  9 &1.81 $\pm$ 1.03E-14 &1& -1.00 $\pm$ 0.23 \\  
5&18:39:27.30, -04:18:50.0 &2.2 &23.0 & 47 $\pm$ 10 &1.59 $\pm$ 1.45E-14 &0.93& -1.00 $\pm$ 0.17 \\
6&18:39:24.20, -04:18:52.8 &3.0 &10.6 & 39 $\pm$  9 &2.28 $\pm$ 1.23E-14 &0.67 & -0.65 $\pm$ 0.48  \\
7&18:39:56.74, -04:15:14.1 &2.4 &12.8 & 39 $\pm$  9 &2.78 $\pm$ 1.49E-14 &0.80 & -0.26 $\pm$ 0.16 \\
8&18:39:11.97, -04:29:02.0 &2.7 &9.1 & 31 $\pm$  8 &1.18 $\pm$ 0.85E-13 &0.10 & 0.002 $\pm$ 0.001 \\   
9&18:40:14.58, -04:16:06.2 &2.4 &12.2 & 40 $\pm$  9 &6.69 $\pm$ 2.99E-14 &0.32 & -0.08 $\pm$ 0.04 \\
10&18:39:29.07, -04:21:48.1 &3.1 &8.3 & 28 $\pm$  8 &1.51 $\pm$ 1.10E-14 &1 & -1.00 $\pm$ 0.25 \\   
11&18:40:09.32, -04:09:33.4 &2.9 &7.4 & 27 $\pm$  7 &2.68 $\pm$ 2.51E-14 &0.48
& -0.62 $\pm$ 0.60 \\
\end{tabular}
\tablecomments{
The likelihood of detection (LoD), counts (Cts), absorbed flux and
 hardness ratio
 (HR) are
from the combined
PN, MOS1 and MOS2 data. The flux is estimaded for the 0.2-12 keV band,
assuming a hydrogen column of $1.5 \times 10^{22}$cm$^{-2}$. Soft fr. is the fraction of the flux in the 0.2-4.5 keV band.
 }
\label{cat1}
\end{table}
}

\clearpage

\rotate{
\begin{table}
\hspace{-3cm}
\scriptsize
\caption[]{Candidate optical counterparts and tentative classification of X-ray sources in the field of SNR G27.8+0.6}
\begin{tabular}{cccccc}
\hline\hline\\
  Src ID  & USNO/2MASS$^{a}$ & Dist$^{b}$ & ${\rm log}(f_{\rm x}/f_{\rm opt})$ & Class & Prob$^{c}$ \\
\hline \\
  & & arcsec & & &  \\
\hline \\
1 & $\cdots$   & $\cdots$  & $>1.3$   &PWN? & $\cdots$ \\
2 &0855-0384773/18401094-0424451 &2.5/2.4 &-3.7 &star  & 0.85/0.86 \\
3 &0856-0387120/18394759-0421233    &5.7/2.2    &-1.7    & AGN/CV & 0.44/0.88   \\
4 &0857-0385700/18395497-0415237 &0.6/0.4 &-1.7 &star & 0.99/0.99 \\
5 &0856-0386213/18392730-0418533 &1.8/3.3 &-1.7 &star  & 0.92/0.76 \\
6 &0856-0386032/18392418-0418497 &3.5/3.1 &-1.6 &star & 0.73/0.78 \\
7 & $\cdots$   & $\cdots$   & $>0.1$   & $\cdots$  & $\cdots$ \\
8 &0855-0381458/18391172-0429001 &4.6/4.1 &-2.4 &star & 0.59/0.65 \\
9 &0857-0386237/18401436-0416068 &3.3/3.3 &-1.3 &star & 0.76/0.76 \\
10 &0856-0386293/18392887-0421468 &3.4/3.1 &-2.0 &star & 0.75/0.78 \\
11 &0858-0387113/18400912-0409349 &1.7/3.4 &-2.4 &star & 0.93/0.75 \\
\\
\hline\\
\end{tabular}
\\$^{a}$ the closest optical(USNO)/NIR(2MASS) source within the 3$\sigma$
error circle of the X-ray position  \\
$^{b}$ the distance between the USNO/2MASS and the X-ray position \\
$^{c}$ the probability that the optical/NIR source is the true counterpart
\label{opt1}

\end{table}
}

\clearpage

\rotate{
\begin{table}
\hspace{-3cm}
\scriptsize
\caption[]{X-ray sources in the field of SNR G28.8+1.5}
\begin{tabular}{cccccccccccc}
\hline\hline\\
  Src  & Position  & Err   & LoD & Cts & Flux & Soft fr. & HR \\
\hline \\
  & J2000   & arcsec   &  &  & erg s$^{-1}$cm$^{-2}$     
   &  &  &  &  & \\
\hline \\
1 &18:39:03.041, -02:42:01.94 &1.5 &1820.0 &1389 $\pm$ 49 &1.72 $\pm$
0.09E-13 &0.88 & -0.91 $\pm$ 0.18 \\
2 &18:39:39.844, -02:49:35.62 &1.6 &579.6 &417 $\pm$ 22 &2.05 $\pm$
0.28E-13 &0.65 & -0.31 $\pm$ 0.04 \\
3 &18:39:47.509, -02:45:05.47 &1.6 &592.7 &417 $\pm$ 22 &1.84 $\pm$
0.38E-13 &0.52 & -0.79 $\pm$ 0.24 \\
4 &18:39:42.684, -02:41:15.97 &1.8 &190.8 &109 $\pm$ 12 &2.61 $\pm$ 1.82E-14 &0.84
& -1.0 $\pm$ 0.1  \\  
5 &18:38:58.595, -02:43:56.28 &1.7 &166.8 &206 $\pm$ 17 &1.23 $\pm$
0.14E-13 &0.29 & 0.48 $\pm$ 0.1 \\
6 &18:39:30.338, -02:43:40.26 &1.7 &149.1 &208 $\pm$ 22 &1.03 $\pm$
0.21E-13 &0.62 & -0.14 $\pm$ 0.03 \\
7 &18:39:19.516, -02:45:41.11 &1.7 &148.4 &173 $\pm$ 16 &1.06 $\pm$
0.15E-13 &0.37 & 0.52 $\pm$ 0.12 \\
8 &18:38:53.640, -02:36:55.80 &1.8 &83.0 &134 $\pm$ 15 &1.46 $\pm$ 0.24E-13 &0.22
& 0.31 $\pm$ 0.07 \\
9 &18:39:56.135, -02:42:06.84 &2.0 &91.1 &129 $\pm$ 16 &2.54 $\pm$ 0.61E-13 & 0.25
& 0.41 $\pm$ 0.12 \\
10 &18:39:32.438, -02:37:10.31 &1.9 &50.5 & 94 $\pm$ 13 &2.08 $\pm$
0.37E-13 &0.14 & 0.84 $\pm$ 0.56 \\ 
11 &18:38:56.855, -02:52:12.36 &2.0 &40.4 & 90 $\pm$ 15 &1.67 $\pm$
0.38E-13 & 0.22 & 0.46
$\pm$ 0.21 \\
12 &18:38:36.507, -02:34:18.19 &1.9 &45.2 & 91 $\pm$ 12 &2.13 $\pm$ 0.98E-14 & 1
& -1.0 $\pm$ 0.17   \\   
13 &18:38:43.052, -02:47:02.80 &1.9 &50.6 &108 $\pm$ 16 &6.74 $\pm$
1.69E-14 &0.36 & 0.26
$\pm$ 0.08 \\
14 &18:38:33.725, -02:48:18.47 &2.2 &21.7 & 73 $\pm$ 13 &6.62 $\pm$
2.14E-14 & 0.25 & 0.30
$\pm$ 0.12 \\
15 &18:39:10.057, -02:46:49.66 &2.0 &22.6 & 72 $\pm$ 13 &1.98 $\pm$
0.76E-14 & 0.52  &
-0.02 $\pm$ 0.01 \\
16 &18:39:38.748, -02:45:13.75 &2.1 &28.2 & 61 $\pm$ 11 &1.55 $\pm$ 1.06E-14 & 1 & -1.0 $\pm$ 0.17  \\   
17 &18:39:44.507, -02:49:22.19 &2.4 &18.6 & 60 $\pm$ 11 &8.04 $\pm$
2.87E-14 & 0.38 & 0.53
$\pm$ 0.26 \\
18 &18:38:56.257, -02:39:04.03 &2.1 &21.1 & 67 $\pm$ 13 &5.12 $\pm$
1.50E-14 &0.25 & 0.43
$\pm$ 0.20 \\
19 &18:38:49.592, -02:39:43.74 &2.0 &23.0 & 73 $\pm$ 13 &7.02 $\pm$
1.63E-14 & 0.16 & 0.46 $\pm$ 0.21 \\
20 &18:38:52.798, -02:53:27.74 &2.1 &19.0 & 54 $\pm$ 10 &1.87 $\pm$ 1.13E-14 & 0.77
& -0.85 $\pm$ 0.17   \\   
21 &18:39:40.212, -02:41:14.89 &2.0 &24.0 & 41 $\pm$ 10 &1.60 $\pm$ 2.19E-14 & 0.30
& -0.51 $\pm$ 0.43   \\   
22 &18:38:32.954, -02:47:15.94 &2.0 &21.4 & 55 $\pm$ 11 &2.55 $\pm$
1.31E-14 &0.70 & 0.02
$\pm$ 0.01 \\
23 &18:39:09.179, -02:48:26.14 &2.3 &17.4 & 64 $\pm$ 12 &4.89 $\pm$
1.31E-14 & 0.23 &
0.29 $\pm$ 0.11 \\
24 &18:38:51.662, -02:50:17.99 &2.4 &13.5 & 39 $\pm$ 10 &6.55 $\pm$
1.89E-14 & 0.14 & 0.66 $\pm$ 0.62 \\
25 &18:39:45.213, -02:32:22.42 &3.3 &8.8 & 35 $\pm$ 12 &1.34 $\pm$ 1.10E-13 &0.30
& -0.06 $\pm$ 0.04  \\  
26 &18:39:20.415, -02:32:08.09 &2.4 &12.6 & 46 $\pm$ 11 &4.43 $\pm$
2.46E-14 & 0.46 & 0.42
$\pm$ 0.23 \\
27 &18:39:33.050, -02:39:41.15 &2.1 &14.4 & 59 $\pm$ 12 &2.67 $\pm$
1.25E-14 & 0.46 & -0.06 $\pm$ 0.03 \\
28 &18:38:26.545, -02:41:03.59 &2.2 &14.0 & 56 $\pm$ 12 &2.42 $\pm$
1.44E-14 & 0.28 & 0.09
$\pm$ 0.04 \\
\end{tabular}
\tablecomments{
The likelihood of detection (LoD), counts (Cts), absorbed flux and
 hardeness ratio
 (HR) are
from the combined
PN, MOS1 and MOS2 data. The flux is estimated for the 0.2-12 keV band, 
assuming a hydrogen column of $2 \times 10^{22}$cm$^{-2}$. Soft fr. is the fraction of the flux in the 0.2-4.5 keV band.
 }
\label{cat2}
\end{table}
}

\clearpage

\rotate{
\begin{table}
\hspace{-3cm}
\scriptsize
\caption[]{Candidate optical counterparts and tentative classification of X-ray sources  in the field of SNR G28.8+1.5}
\begin{tabular}{cccccc}
\hline\hline\\
  Src  & USNO/2MASS$^a$ & Dist$^b$ & ${\rm log}(f_{\rm x}/f_{\rm opt})$ & Class & Prob$^c$  \\
\hline \\
  & & arcsec & & &  \\
\hline \\
1 &0872-0524632/18390292-0242033 &3.4/2.2 &-0.9 &star & 0.76/0.89 \\
2 &0871-0519069/18393984-0249372 &1.0/1.6 &-0.8 &star & 0.98/0.94 \\
3 &0872-0528442/18394750-0245056 &1.0/0.2 &-2.1 &star & 0.98/0.99 \\
4 &0873-0529676/$\cdots$ &0.3/$\cdots$ &-0.8 &star & 0.99/$\cdots$ \\
5 &0872-0524304/18385867-0243556 &2.9/1.3 &-0.5 &AGN/CV & 0.82/0.96  \\
6 &0872-0526884/18393041-0243446 &5.6/4.5 &-0.2 &AGN/CV & 0.47/0.61 \\
7 &0872-0525933/18391953-0245476 &2.8/6.6 &-0.9 &AGN/CV & 0.83/0.35 \\
8 &0873-0525927/18385353-0237012 &4.8/5.7 &-1.2 &AGN/CV & 0.57/0.45 \\
9 &0872-0529117/18395631-0242087 &4.2/3.3 &-1.2 &AGN/CV &0.65/0.77 \\
10 &0873-0528766/18393230-0237074&3.6/3.5 &-1.1 &AGN/CV & 0.73/0.74 \\
11 &0871-0515153/18385665-0252137 &2.8/3.2 &-1.2 &AGN/CV & 0.83/0.78 \\
12 &0874-0529861/18383663-0234189 &1.8/2.1 &-3.8 &star & 0.92/0.90 \\
13 &0872-0523159/18384311-0247016 &0.5/1.5 &-1.3 &star or GAL & 0.99/0.95 \\
14 &0871-0513364/18383390-0248209 &3.4/3.7 &0.0 &AGN/CV & 0.76/0.72 \\
15 &0872-0525201/18390987-0246532 &1.6/4.5 &-1.5 &star & 0.94/0.61  \\
16 &0872-0527751/18393871-0245156 &1.0/2.0 &-1.7 &star &0.98/0.91  \\
17 &0871-0519498/18394421-0249239 &5.0/4.7 &-0.8 &AGN/CV & 0.55/0.59 \\
18 &0873-0526103/18385657-0239038 &4.7/4.8 &-0.8 &AGN/CV & 0.59/0.57 \\
19 &0873-0525656/$\cdots$ &3.9/$\cdots$ &-1.3 &AGN/CV & 0.69/$\cdots$ \\
20 &0871-0514833/18385276-0253274 &0.5/0.5 &-1.4 &star & 0.99/0.99 \\
21 &0873-0529499/18394054-0241153 &4.4/5.0 &-2.5 &star & 0.63/0.55 \\
22 &0872-0522454/$\cdots$ &5.6/$\cdots$ &-0.7 &AGN/CV & 0.47/$\cdots$ \\
23 &0871-0516197/18390912-0248290 &0.3/3.0 &-2.0 & star or GAL &0.99/0.80 \\
24 &0871-0514762/18385212-0250137 &2.0/8.1 &-1.4 &AGN/CV &0.91/0.20 \\
25 &0874-0534435/18394499-0232148 &2.0/8.2 &-0.6 & star & 0.91/0.20 \\
26 &0874-0532767/18392034-0232082 &1.3/1.1 &-0.9 &AGN/CV & 0.96/0.97 \\
27 &0873-0528856/18393299-0239382 &2.1/3.0 &-1.0 &star & 0.90/0.80 \\
28 &0873-0524047/18382638-0241016 &3.2/3.0 &-1.6 &star or GAL & 0.78/0.80 \\
\\
\hline\\
\end{tabular}
\\$^{a}$ the closest optical(USNO)/NIR(2MASS) source within the 3$\sigma$
error circle of the X-ray position  \\
$^{b}$ the distance between the USNO/2MASS and the X-ray position \\
$^{c}$ the probability that the optical/NIR source is the true counterpart
\label{opt2}
\end{table}
}

\clearpage

\bibliographystyle{apj}
\bibliography{./report, ./paper_ref, ./add}

\end{document}